\documentclass[12pt,a4]{article}

\usepackage{amsmath}
\usepackage{amssymb}
\usepackage{epsfig}
\usepackage{graphicx}
\usepackage{mcite}

\makeatletter
\renewcommand{\fnum@table}{\textbf{\tablename~\thetable}}
\renewcommand{\fnum@figure}{\textbf{\figurename~\thefigure}}
\makeatother

\newcounter{myenumi}

\renewcommand{\themyenumi}{\roman{myenumi}}
{\end{list}}

\newlength{\myem}
\newcounter{mysubequation}[equation]

\makeatletter
\renewcommand{\section}{\@startsection{section}{1}{0em}{-\baselineskip}%
{\baselineskip}{\normalfont\large\bfseries}}
\renewcommand{\subsection}%
{\@startsection{subsection}{2}{0em}{-0.7\baselineskip}%
{0.7\baselineskip}{\normalfont\bfseries}}
\makeatother

\newcommand{\SM}{\mathrm{SM}}
\newcommand{\bulk}{\mathrm{bulk}}
\newcommand{\brane}{\mathrm{brane}}
\newcommand{\Lag}{{\cal L}}
\newcommand{\hc}{\mathrm{h.c.}}

\newcommand{\eps}{\varepsilon}
\newcommand{\UF}{U(1)_F}

\newcommand{\diag}{\mathrm{diag}}

\begin{document}

\begin{titlepage}

\renewcommand{\thefootnote}{\alph{footnote}}

\vspace*{-3.cm}
\begin{flushright}
TUM-HEP-624/06
\end{flushright}

\vspace*{0.5cm}

\renewcommand{\thefootnote}{\fnsymbol{footnote}}
\setcounter{footnote}{-1}

{\begin{center}
{\Large\bf Family Symmetry and Single Right-Handed Neutrino Dominance in Five Dimensions}
\end{center}}

\vspace*{.8cm}
{\begin{center} {\large{\sc
                Marc-Thomas Eisele\footnote[1]{\makebox[1.cm]{Email:}
                eisele@ph.tum.de},~
                Naoyuki Haba\footnote[2]{\makebox[1.cm]{Email:}
                haba@ph.tum.de, on leave of absence from \\
				\makebox[11ex]{} Institute of Theoretical Physics, 
				University of Tokushima, 770-8502, Japan}
                                }}
\end{center}}
\vspace*{0cm}
{\it
\begin{center}

\vspace*{1mm}

       Physik--Department, Technische Universit\"at M\"unchen, \\
       James--Franck--Strasse, 85748 Garching, Germany
\vspace*{1mm}


\vspace*{1cm}

\today
\end{center}}

\vspace*{0.3cm}

\begin{abstract}
We consider several neutrino mass models in an extra-dimensional setting on a quantitative level. All the models are set in a five-dimensional scenario, with the standard model (SM) particles living on a brane, while three additional SM gauge singlets live in the bulk of an extra dimension, which is compactified on a $S^1/Z_2$ orbi\-fold. The spontaneous breaking of an additional, continuous $U(1)$ family symmetry is used to generate suitable neutrino mass matrices via single right-handed neutrino dominance through the corresponding five-dimensional extension of the see-saw mechanism. In this manner possible problems of this combination for some models in four dimensions could be overcome. The considered models differ with respect to the charges under the family symmetry and the nature of the five-dimensional Majorana mass term.

\end{abstract}

\vspace*{.5cm}

\end{titlepage}

\newpage

\renewcommand{\thefootnote}{\arabic{footnote}}
\setcounter{footnote}{0}


\section{Introduction}
The data on the various neutrino masses and mixing angles has become more and more precise. Current bounds on the $3\sigma$ level from Ref.\cite{Maltoni:2004ei} are
(see also Refs.\cite{Fogli:2003th, *Bahcall:2004ut,*Bandyopadhyay:2004da})
\begin{equation}\label{eq obs values}
	\begin{array}{lcl}
		\Delta m^2_{21} &=&(7.1-8.9)\cdot 10^{-5}\mathrm{eV}^2 \, ,\\
		\Delta m^2_{31} &=&(1.4-3.3)\cdot 10^{-3}\mathrm{eV}^2 \, ,\\
		\sin ^2 \theta _{12} &=& 0.23-0.38 \, ,\\
		\sin ^2 \theta _{23} &=& 0.34-0.68 \, ,\\
		\sin ^2 \theta _{13} &<& 0.051 \, .
		\end{array}
	\end{equation}
	
However, the origin of these values is still a mystery. While it is usually assumed, that the so-called see-saw mechanism \cite{Minkowski:1977sc,*Yanagida:1979as,*Gell-Mann:1980vs} is responsible for the small scale of the neutrino masses, there are many more competing theories that try to explain the mixing of the light neutrinos and possible orderings of their masses.
In particular, in the context of extra dimensions and neutrino masses and mixings, there has been a considerable scientific output, e.g. Refs. \cite{Arkani-Hamed:1998vp,Dienes:1998sb,Lukas:2000wn,Lukas:2000rg,Faraggi:1999bm,*Dvali:1999cn,*Das:1999dx,*Mohapatra:1999zd,*Ioannisian:1999cw,*Mohapatra:1999af,*Ioannisian:1999sw,*Barbieri:2000mg,*Lukas:2000fy,*Ma:2000wp,*Mohapatra:2000wn,*Ma:2000gf,*Dienes:2000ph,*Mohapatra:2000px,*McLaughlin:2000zf,*Ioannisian:2000en,*Agashe:2000rw,*Cosme:2000ib,*Caldwell:2000zn,*Abazajian:2000hw,*Caldwell:2001dj,*Lam:2001gy,*DeGouvea:2001mz,*Frere:2001ug,*Lam:2001iy,*Davoudiasl:2002fq,*Mohapatra:2002ug,*Kim:2002im,*Cao:2003yx,*Hewett:2004py,*Cao:2004tu,*Dudas:2005vn,Haba:2006dz}.

A possible explanation for a hierarchy in the light neutrino sector is given by the so called single right-handed neutrino dominance (SRND) \cite{King:1998jw,*Davidson:1998bi,*King:2003jb,King:1999cm,King:1999mb}, where a mass hierarchy in the right-handed neutrino sector can lead to an effective Majorana mass matrix  for the light neutrinos which also yields a hierarchy. Further, an additional hierarchy in the Yukawa couplings of left- and right handed neutrinos can lead to a large $\theta _{23}$ and a very small $\theta _{13}$. Of course, the contribution of the two heavier right-handed neutrinos should also not be completely negligible, since this would predict only one massive, light neutrino, in contradiction with observation.

One way to account for such hierarchies is through an additional family symmetry $U(1)_F$, under which the right- and left-handed neutrinos have different charges in each generation. This might naturally lead to a mass hierarchy in the right-handed neutrino sector as well as to the mentioned hierarchy in the Yukawa couplings of the standard model (SM) leptons to each right-handed neutrino. However, in many models (i.e. for many charge assignments) it might happen that the hierarchies in the heavy Majorana matrix and the Yukawa couplings tend to annihilate each other and hereby disfavor a SRND.%
\footnote{Of course, one should admit that, due to the weakness of the hierarchy in the neutrino sector, SRND can still be possible for these cases without dramatic tuning.} %
This possible annihilation might be overcome by means of the five-dimensional see-saw mechanism, as we illustrate in this paper.

In this context, we treat several models that were introduced in a corresponding letter \cite{Haba:2006gt} on a quantitative basis and generalize their charge assignments under an additional continuous family symmetry. All the models make use of the altered see-saw mechanism in the presence of a fifth dimension to establish a SRND. %
In this setting the presence of an extra dimension can vary the $1/M$ term present in the four-dimensional see-saw formula \cite{Dienes:1998sb,Lukas:2000wn,Lukas:2000rg}. As this paper illustrates, this might alter the phenomenology of the considered models significantly. In particular, this might also be the reason why a possible annihilation of the hierarchies in the Dirac and Majorana sector for some models might be prevented by the presence of a fifth dimension, as mentioned before.

The structure of the paper is as follows: After a brief review of SRND in section two, we outline the see-saw mechanism in five dimensions in section three (as introduced in Refs. \cite{Dienes:1998sb,Lukas:2000wn,Lukas:2000rg}). This is followed by an introduction to possible mass-term hierarchies due to a $U(1)_F$ flavor symmetry in section four. Section five comprises the main part of this paper, where we consider several neutrino mass models with the features described earlier in this section. During the treatment of the 
 first model the basic ideas behind the models will become more evident. We also determine possible parameter values that lead to the observed values for neutrino masses and mixings for this model under the corresponding assumptions and derive guidelines for finding further ones. However, some additional tuning in the heavy sector of this model might be necessary to justify some of the corresponding assumptions. This seems less of a problem in the following models. However, we also use parts of the analysis of the first model at later points. In the first two models, the Majorana mass term of the SM singlets will be a five-dimensional, vectorlike mass term, leading to a cotangent suppression for the corresponding see-saw mechanism. This vectorlike mass term violates Lorentz invariance in the additional dimension. Though this can be possible, it might be considered as unnatural by the reader. This is not the case for the two further models we consider, where the Majorana mass term of the SM singlets will instead be a five-dimensional, scalar mass term. In this case Lorentz invariance is conserved in all five dimensions. The now induced hyperbolic cotangent suppression leads to a slightly different analysis as for the first two cases.  For all models we present parameter sets that yield suitable neutrino masses and mixings under the respective assumptions, while we also give approximate formulae and guidelines for finding further sets. After the analysis of the models we conclude in section six. A short review on mixing angles due to mass mixings and some of the more complex formulae that can help finding suitable parameter sets for the various models are presented in the appendix.


\section{Single Right-Handed Neutrino Dominance}

In this section we briefly summarize the SRND mechanism as presented in Refs.\cite{King:1998jw,*Davidson:1998bi,*King:2003jb,King:1999cm,King:1999mb}.

The key idea can already be seen in the case of two generations, which will later be the 23-sector. In this case we can write the right-handed Majorana mass matrix $M_N$, which can always be diagonalized, and the Dirac mass matrix $m_D$ as
\begin{equation}
	M_N \equiv \left( \begin{array}{cc} M_1&0\\ 0&M_2 \end{array}\right)
	\quad \mbox{and} \quad
	m_D \equiv \left( \begin{array}{cc} a&b\\ c&d \end{array}\right) \, .
	\end{equation}
The see-saw formula
\begin{equation}
\label{eq 4d see-saw}
m_{LL} = m_D M^{-1} m_D^T
\end{equation}
then yields the effective Majorana mass matrix for the light neutrinos
\begin{equation}
	m_{LL}\equiv \left( \begin{array}{cc} 
	\frac {a^2}{M_1}+\frac {b^2}{M_2}&\frac {ac}{M_1}+\frac {bd}{M_2}\\
	\frac {ac}{M_1}+\frac {bd}{M_2}&\frac {c^2}{M_1}+\frac {d^2}{M_2}
	 \end{array}\right) \, .
	\end{equation}
If there is no hierarchy in the Dirac masses, we can already see, that all entries of $m_{LL}$ are of the same order of magnitude, which naturally leads to large mixing of the two considered states. Additionally, if there is a hierarchy in the heavy sector (e.g. $M_1\ll M_2$), the rank of this matrix reduces approximately to one, yielding a hierarchy in the effective light masses. In the case of $M_1\ll M_2$ the eigenvector with the heavy mass is $(a,c)^T$. In this case the tangent of the mixing angle is $a/c$ (compare eq.(\ref{eq tan23})). If we extend the model to three generations we see, that the new Yukawa coupling to the lightest of the heavy neutrinos has to be somewhat smaller than $a$ and $c$ to yield a small $\theta _{13}$ (compare eq.(\ref{eq tan13})). This can ,e.g., be achieved with a suitable family symmetry, which we will discuss at a later point.

However, a complete dominance of one of the right-handed neutrinos is not realized in nature, since in this case the rank of $m_{LL}$ would be one, yielding two massless light neutrinos, in contrast to observation. Therefore, the contribution of further heavy neutrinos should not be completely neglected.


\section{Five-Dimensional See-Saw}\label{sec five d see saw}

This section mainly summarizes the see-saw mechanism in the five-dimensional setting used in this paper as it has been developed in Refs. \cite{Arkani-Hamed:1998vp,Dienes:1998sb,Lukas:2000wn,Lukas:2000rg}. 

For the whole paper we work in the representation
\begin{equation}
	\gamma _\mu = \left( \begin{array}{cc} 0& \sigma _\mu\\
		\bar \sigma _\mu &0 \end{array} \right) \, ,\quad
	\gamma _5=i\gamma_4 = \left( \begin{array}{cc} 1_{2\times 2}& 0\\
		0&-1_{2\times 2} \end{array} \right) \, ,
	\end{equation}
where $\sigma _\mu=(1_{2\times 2}, -\sigma _i)$ 
and $\bar \sigma _\mu=(1_{2\times 2}, \sigma _i)$ 
with the Pauli matrices $\sigma _i$.

We consider a five-dimensional scenario, where the SM particles live on a four-dimensional brane. Their behavior is determined by the usual four-dimensional Lagrangian density of the SM $\Lag_\SM$
\begin{equation}
	S_\SM=\int d^4x\,dy\, {\cal L}_{SM}\,\delta (y)
	\end{equation}
and possible couplings to particles living in the bulk.

We now introduce three additional fields $\Psi _i$ living in an extra dimension compactified on an orbifold $S^1/Z_2$ with radius $R$.
We also define the corresponding charge conjugate fields $\Psi ^c_i$ by
\begin{equation}
	\Psi ^c _i \equiv \left( \begin{array}{cc}
			0&\epsilon \\ \epsilon&0 \end{array} \right)
			\Psi_i^* \, ,
	\end{equation}
where $\epsilon=i\sigma _2$.

Under the parity transformation $P_5: y \rightarrow -y$ the $\Psi _i$ transform as 
\begin{equation}\label{eq psi transf}
	P_5 \Psi _i= \gamma _5 \Psi _i \, .
	\end{equation}
Neglecting their couplings to the SM for the moment, their most general behavior can then be described by the bulk action
\begin{equation}\label{eq bulk action}
	S_\bulk=\int d^4x\,dy\, \left[ 
		\overline{\Psi}_i i\gamma ^\alpha \partial _\alpha \Psi _i
		-\frac 12 (M^S_{ij} \overline{\Psi ^c}_i \Psi _j
		+M^V_{ij} \overline{\Psi ^c}_i \gamma _5 \Psi _j
		+\hc )\right] \, ,
	\end{equation}
with the scalarlike Majorana mass $M^S$ and vectorlike Majorana mass $M^V$.%
\footnote{We write $S$ and $V$ as sub- or superscripts according to convenience in each case.}
Additionally, one can allow Yukawa coupling terms between $\Psi$ and the SM
\begin{equation}\label{eq 5d Yuk int}
	S_\brane=\int d^4x\,dy\, \left[
		- \frac {g_{ij}}{\sqrt{M_5}}
		\overline{\Psi}_i P_L H 
		\left( (\epsilon _{2}\ell_j)^T,0,0 \right)^T  + \hc \right]
		\delta (y)+S_\SM  \,  \, ,
	\end{equation}
where $M_5$ is an additional mass term, which is needed for a dimensionless action, $\ell _j$ are the
three left-handed SM lepton doublets represented by two-dimensional
Weyl spinors, $H$ is the SM Higgs doublet, $g_{ij}$ are Yukawa
couplings, $P_L\equiv(1-\gamma
_5)/2$, and $\epsilon _2\equiv\epsilon$ but now acting on the $SU(2)$ doublet $\ell$. (A
similar Yukawa term for $\Psi ^C$ is forbidden by parity $P_5$.)
 The Higgs boson component that develops a vacuum expectation
value (VEV) $v$ now yields Dirac mass terms for the left-handed SM
neutrinos $\nu _j$ and the right-handed components of $\Psi _i$. 
In our models $M_5$ will be the suppression scale that leads to small neutrino masses, therefore we assume it to be rather heavy. Its actual size will also depend on the size of the various Yukawa couplings, which will be determined by a family symmetry in the considered models. If one wishes to make the considered models more restrictive, one can additionally assume $M_5$ to be the five-dimensional Planck scale, in which case $M_5$ and the four-dimensional Planck scale $M_4$ are typically taken to
be related
through the radius $R$ of the extra dimension by $M_5^3 R\approx M_4^2$ \cite{Arkani-Hamed:1998rs,*Antoniadis:1998ig}.

The above transformation rules (eq.(\ref{eq psi transf})) allow us to write 
\begin{equation}
	\Psi _i (x,y)= \frac 1{\sqrt{\pi R}}
	\left( \begin{array}{r}
	\frac 1{\sqrt{2}} \Psi ^{(0)}_{R,i}(x) + \sum \limits^\infty _{n=1}
		\cos (ny/R) \, \Psi ^{(n)}_{R,i}(x) \\
		\sum \limits^\infty _{n=1}
		\sin (ny/R)\,  \Psi ^{(n)}_{L,i}(x) \end{array}
		\right),
	\end{equation}
where $\Psi ^{(n)}_{R/L,i}(x)$ are right- and left-handed Weyl spinors, respectively.

Integrating over the fifth dimension, we can now write the complete
four-dimensional action responsible for neutrino masses as
\begin{eqnarray}
S=&&{\displaystyle \int} d^4x \, \Bigg[ 
	\sum \limits_{n=0}^\infty \hat \Psi^{c(n)\dagger}_{R,i}
	i \sigma ^\mu \partial _\mu  \hat \Psi^{c(n)}_{R,i}
	+\sum \limits_{n=1}^\infty \Psi ^{(n)\dagger}_{L,i}
	i \sigma ^\mu \partial _\mu \Psi ^{(n)}_{L,i}
	+ \Bigg( \sum \limits_{n=1}^\infty \frac nR \hat \Psi^{c(n)T}_{R,i} 
	\epsilon \Psi ^{(n)}_{L,i} \nonumber \\
	&&
	+ \frac {g_{ij} v}{\sqrt{2\pi R M_5}} \hat \Psi^{c(0)T}_{R,i}
	\epsilon \, \nu _j
	+ \sum \limits_{n=1}^\infty \frac {g_{ij} v}{\sqrt{\pi R M_5}} 
	\hat \Psi^{c(n)T}_{R,i}	\epsilon \,\nu _j 
	-\sum \limits_{n=0}^\infty \frac 12 (M_{S,ij}^*+M_{V,ij}^*)
	\hat \Psi^{c(n)T}_{R,i} \epsilon \hat \Psi ^{c(n)}_{R,j}
	\nonumber \\ && 
	-\sum \limits_{n=1}^\infty \frac 12 (-M_{S,ij}+M_{V,ij})
	\Psi^{(n)T}_{L,i} \epsilon \Psi ^{(n)}_{L,j}
	+\hc \Bigg) \Bigg]+S_{\SM} \,  \,  ,
	\end{eqnarray}
where $\hat \Psi^{c(k)}_{R,i}\equiv -\epsilon (\Psi ^{(k)}_{R,i})^*$ is the left-handed Weyl spinor corresponding to the right-handed $\Psi ^{(k)}_{R,i}$.

We may combine the various mass terms to one large Majorana mass matrix. In the completely left-handed basis $(\nu,\hat \Psi ^{c(0)}_{R,i},\dots,\Psi ^{(n)}_{L,i},
\hat \Psi ^{c(n)}_{R,i},\dots)$ we can now write
\begin{equation}\label{eq mass matrix gen}
	\left( \begin{array}{cccccc}
	0& - m_D/\sqrt{2} &\cdots& 0& - m_D&\cdots \\
	- m_D^T/\sqrt{2}& M_S^*+M_V^*&\cdots& 0& 0&\cdots \\
	\vdots&\vdots&\ddots&\vdots&\vdots&\ddots \\
	0&0&\cdots& -M_S+M_V&- n/R&\cdots \\
	- m_D^T&0&\cdots&-n/R& M_S^*+M_V^*&\cdots \\
	\vdots&\vdots&\ddots&\vdots&\vdots&\ddots 
	\end{array} \right) \, ,
	\end{equation}
with the Dirac mass $m_{D,ij} \equiv g_{ji}v/\sqrt{\pi R M_5}$, where we switched the indices in this definition for later convenience.

For arbitrary parameter values it cannot easily be seen how to diagonalize this matrix. However, in the general case of arbitrary $M_S$ and $M_V$ one can still make a perturbative expansion for small $m_D$ in which case one can come up with a modified see-saw formula:

For a fixed pair $\Psi ^{(n)}_{L,i},\hat \Psi ^{c(n)}_{R,j}$ we can define the Majorana sub-matrix
\begin{equation}
	M_n\equiv \left( \begin{array}{cc}
		 -M_S+M_V&- n/R\\
		 -n/R& M_S^*+M_V^* \end{array} \right) \, .
	\end{equation}
Next we define the small perturbation parameters
\begin{equation}
	\eta _0 \equiv -M_0^{-1}\, \frac{m_D^T}{\sqrt{2}},\quad 
	\eta _n \equiv -M_n^{-1} \left( \begin{array}{c} 0\\m_D^T \end{array}
	\right) \, ,
	\end{equation}
with $M_0\equiv (M_{n})_{22}$, neglecting flavor indices for the moment.

However, since we will sum over an infinite amount of states we also require this sum to be small. Therefore, we define
\begin{equation}
	\eta ^2 \equiv ||\eta _0||^2
	+\sum \limits_{n=1}^\infty ||\eta _n||^2 \, ,
	\end{equation}
with $|| \eta_0 ||\equiv \sum_{i,j} |\eta _{0,ij}|^2$ and the corresponding definition for $|| \eta _n||$, while we also demand $\eta ^2 \ll 1$ in addition to $|\eta _i|\ll 1$.

If this condition is fulfilled, we can make a change of basis defined by
\begin{eqnarray}
	\hat \Psi ^{c(0)'}_{R} &\equiv& 
		\hat \Psi ^{c(0)}_{R} +\eta _0 \nu \,  ,\\
	\left( \begin{array}{c}\Psi ^{(n)'}_{L} \\
		\hat \Psi ^{c(n)'}_{R} \end{array} \right) &\equiv&
		\left( \begin{array}{c} \Psi ^{(n)}_{L}\\
		\hat \Psi ^{c(n)}_{R} \end{array} \right)
		+\eta _n \nu  \, , \\
	\nu ' &\equiv& \nu - \eta ^\dagger _0 \hat \Psi ^{c(0)}_{R}
		- \sum \limits _{n=1}^\infty \eta ^\dagger _n
		\left( \begin{array}{c}\Psi ^{(n)}_{L}  \\
		\hat \Psi ^{c(n)}_{R}\end{array} \right) \, ,
	\end{eqnarray}	
with implicit flavor indices.

Then, the mass part of the Lagrangian is to respective leading order
\begin{equation}
	\Lag _{\mathrm{mass}}=
	-\frac 12 \nu'^T m' \epsilon \nu'
	-\frac 12 \hat \Psi ^{c(0)'T}_{R} M_0 \epsilon 
		\hat \Psi ^{c(0)'}_{R}
	-\frac 12 \sum \limits _{n=1}^\infty 
	\left( \begin{array}{c} \Psi ^{(n)'}_{L}\\
		 \hat \Psi ^{c(n)'}_{R}\end{array} \right) ^T
		M_n 
		\left( \begin{array}{c} \epsilon \Psi ^{(n)'}_{L} \\
		\epsilon \hat \Psi ^{c(n)'}_{R} \end{array} \right)
	+ \hc  \, ,
	\end{equation}
where $\epsilon$ acts on the spinor space and
\begin{equation} \label{eq 5d see-saw gen}
	m'\equiv - \frac 12 m_D M_0^{-1}m_D^T-\sum \limits_{n=1}^\infty 
	(0,\,m_D)\, M_n^{-1}
	\left( \begin{array}{c} 0\\m_D^T \end{array} \right) \, .
	\end{equation}
If one can simultaneously diagonalize $M_S$ and $M_V$, one can also unitarily transform $\hat \Psi ^{c(n)'}_{R,j}$ and $\Psi ^{(n)'}_{L,i}$ in such a way, that the block matrices of $M_n$ on the main diagonal become diagonal themselves, without changing the off-diagonal block-matrices. In this case
eq.(\ref{eq 5d see-saw gen}) can be simplified further:	
\begin{equation}\label{eq 5D see-saw}
	m'=-m_DU^* \frac {\pi R}{2} 
	\left( \begin{array}{ccc} 
		\ddots&0&0 \\
		0&\sqrt{\frac{M^V_i-M^S_i}{M^{V*}_i+M^{S*}_i}}
			\cot(\pi R\
			\sqrt{(M^V_i-M^S_i)(M^{V*}_i+M^{S*}_i)})&0 \\
		0&0&\ddots
		\end{array} \right)
		U^\dagger m_D^T \, , 
	\end{equation} 
where the non-zero entries depend on the row number $i$ and $U^T M_S U \equiv \diag (M^S_1,M^S_2,M^S_3)$ as well as $U^T M_V U \equiv \diag (M^V_1,M^V_2,M^V_3)$.%
\footnote{The complex square roots $\sqrt{(M^V_i-M^S_i)}$ and $\sqrt{(M^V_i+M^S_i)^*}$ of course can take two possible values. However, in this case it is only important to use the same value at the respective two positions, where these terms show up in the above equation.}%

In case $M_V=0$ all the $M^S_i$ can be made real and one can further simplify to
\begin{equation}\label{eq coth see-saw}
	m'=-m_DU^* \frac {\pi R}{2} 
	\left( \begin{array}{ccc} 
		\coth(\pi M_1 R)&0&0 \\
		0&\coth(\pi M_2 R)&0 \\
		0&0&\coth(\pi M_3 R)
		\end{array} \right)
		U^\dagger m_D^T \, , 
	\end{equation}  
while one finds for $M_S=0$
\begin{equation}\label{eq cot see-saw}
	m'=-m_DU^* \frac {\pi R}{2} 
	\left( \begin{array}{ccc} 
		\cot(\pi M_1 R)&0&0 \\
		0&\cot(\pi M_2 R)&0 \\
		0&0&\cot(\pi M_3 R)
		\end{array} \right)
		U^\dagger m_D^T \, , 
	\end{equation}  
where one can again choose $U$ in a way that all $M^V_i$ are real.

\begin{figure}
	\begin{center}\includegraphics[%
	  clip,
	  width=0.5\columnwidth,
	  keepaspectratio]{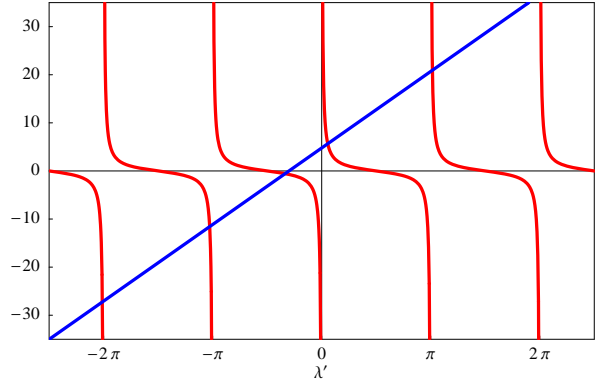}\vspace{-0.5cm}\end{center}
	\caption{\label{fig cot inter}\textit{\small The graphic solution to the 	eigenvalue problem from the text. Each intersection of the straight
	 line and the various cotangent functions represents a solution to the
	 eigenvalue problem represented by eq.(\ref{eq char ev eq}). For this
	  plot we used $m_DR=0.2$ and $M_VR=0.3$.}}
	\end{figure}

In the case, where one can simultaneously diagonalize $m_D$ and $M_{V}$ one can even reduce the problem to a one generation problem and give the exact analytic solution. To see this we go back to eq.(\ref{eq mass matrix gen}). We set $M_S=0$ and $\mathrm{Im} (M_V)=0$ and make a change of basis  for $n\ge 1$ by setting
\begin{equation}
	\left( \begin{array}{c} \Psi^{(n)}_+ \\ \Psi^{(n)} _- 
		\end{array} \right)
	\equiv
	\frac 1{\sqrt{2}}
	\left( \begin{array}{cc} 1&-1 \\ 1&1 \end{array} \right)
	\left( \begin{array}{c} \Psi^{(n)}_{L} 
		\\ \hat \Psi ^{c(n)}_{R}
	 \end{array} \right) \, .
	 \end{equation}
In the new basis $(\nu,\hat \Psi ^{c(0)}_{R},\dots,\Psi ^{(n)}_+,
\Psi ^{(n)}_-,\dots)$ we have
\begin{equation}
	\left( \begin{array}{cccccc}
	0& - m_D/\sqrt{2} &\cdots& m_D/\sqrt{2}&- m_D/\sqrt{2}&\cdots \\
	- m_D/\sqrt{2}& M_V&\cdots& 0& 0&\cdots \\
	\vdots&\vdots&\ddots&\vdots&\vdots&\ddots \\
	m_D/\sqrt{2}&0&\cdots&M_V+n/R&0&\cdots \\
	-m_D/\sqrt{2}&0&\cdots&0&M_V-n/R&\cdots \\
	\vdots&\vdots&\ddots&\vdots&\vdots&\ddots 
	\end{array} \right) \, ,
	\end{equation}
and we remind the reader that we only work with one generation for the moment.
The characteristic eigenvalue equation for the above matrix is
\begin{equation}\label{eq char ev eq}
	\left[ \prod \limits_{k=1}^\infty 
		\left( (M_V-\lambda)^2-\frac {k^2}{R^2}\right) \right]
	\times
	\left[ \lambda(M_V-\lambda)-m_D^2/2+(M_V-\lambda)^2m_D^2
		\sum \limits_{k=1}^\infty
		\frac 1{k^2-(M_V-\lambda)^2R^2}
	\right] .
	\end{equation}
Since every zero of the left bracket is canceled by a singularity in the right one, we only have to look for the zeros of the right bracket.
This yields the transcendental equation
\begin{equation}
	\lambda R=\frac {\pi}2 (m_D R)^2 \cot[\pi R(\lambda -M_V)] \, .
	\end{equation}
The graphical solution of this problem can be seen in fig.\ref{fig cot inter},
 where we make the substitution $\lambda ' \equiv \pi(\lambda-M)R$. Now, the solutions are the crossing points between $\cot(\lambda ')$ and the straight line given by 
\begin{equation}
	\lambda ' \rightarrow \frac 2{(\pi m_D R)^2} \lambda'
		+\frac {2M_V}{\pi m_D^2 R} \, .
	\end{equation}
We can see that the intersection of this straight line and the x-axis is always
at $\lambda_0'=-\pi MR$. Hence, for $MR=1/2$ there is an intersection of this straight line and the cotangent function at $\lambda'=-\pi/2$, which yields $\lambda=0$. Thus, in the case $MR=1/2$	the light neutrino is massless to all orders in $mR$.

In the case of three generations and arbitrary $m_D$ and $M_V$ this simple approach does not work and we have to with the perturbative approach. For the remainder of this paper we therefore require the various $M_{V/S}$ and $1/R$ to be large enough (compared to $m_D$), such that we can neglect higher order terms. This condition also yields an upper limit for the absolute value of the various family charges in the considered models (typically at very large values).


\section{Family Symmetry}

To give the Majorana mass matrix of the additional SM gauge singlets as well the Dirac mass matrix
of the neutrinos a certain structure, one can introduce an additional broken family
symmetry $U(1)_F$, as it was suggested in Ref. \cite{Froggatt:1978nt} (in four dimensions), where the corresponding transformation
properties of a particle also depends on its generation index.
In the context of neutrino masses this has been done many times in the literature, see e.g. Refs.
\cite{King:1999cm,King:1999mb,Ellis:1998nk,*Lola:1999un,*Leontaris:1995be,*Dreiner:1994ra,*Elwood:1998kf,*Irges:1998ax,*Altarelli:1998sr,*Altarelli:1998ns,*Altarelli:1999wi}.

For the considered models we do not specify the nature (global or local) of this
  symmetry. We only assume that emerging side-effects are under
  control, when this symmetry is spontaneously broken. In case of a global symmetry this means that couplings of
  the arising Goldstone boson (familon) must be negligible (see
  e.g. Refs. \cite{Wilczek:1982rv,*Reiss:1982sq,*Gelmini:1982zz,*Feng:1997tn,*Ammar:2001gi,*Diaz-Cruz:2004sp}), while in the case of a local symmetry we assume that possibly emerging
  anomalies are canceled (e.g. by additional heavy particles).

In general, this symmetry (if unbroken) will prevent bare Majorana mass terms for particles that transform non-trivially under it. The same holds for Dirac mass terms, if the left- and right-handed components of a Dirac spinor transform differently under this symmetry.

Let us also introduce an additional scalar particle $\phi$ in the bulk with charge $q^\phi=-1$ under the new symmetry. Higher order effects (e.g. due to further exotic particles) might now introduce Majorana as well as Dirac mass terms for the particles. Let us consider this more closely for the neutrino sector:

We first take a look at eq.(\ref{eq bulk action}) and see that these mass terms are in general forbidden if the $\Psi _i$ carry the respective, non-zero charges $q^\Psi_i$ under $U(1)_F$. However, higher order effects might introduce effective coupling terms as
\begin{equation}\label{eq majaorana mass u(1)_F}
	-\frac 12 \Bigg(\frac {\phi^* }
        {M_5^{3/2}}
         \Bigg)
        ^{q^{\Psi^c}_i+q^{\Psi^c}_j}
        (M^S_{ij} \overline{\Psi ^c}_i \Psi _j
        +M^V_{ij} \overline{\Psi ^c}_i \gamma _5 \Psi _j)
        +\hc \, ,
	\end{equation}
where $q^{\Psi^c}_i = -q^{\Psi}_i$. Also, we implicitly assumed, that all these effective terms are suppressed by powers of
$M_5$.%
\footnote{In principle, new effects could also show up at a mass scale different from $M_5$. However, for reasons of simplicity we will assume
$M_5$ to be the characteristic scale of the effects responsible for the considered mass terms.}

We further assume that $\phi$ develops a VEV $\langle \phi \rangle$, which then leads to Majorana masses for the various $\Psi_i$ as we can see from the above equation.
Let us define 
\begin{equation}\label{eq def eps}
	\frac{\langle \phi \rangle}{M_5^{3/2}} \equiv \eps \, ,
	\end{equation}	 
and require $\eps$ to be smaller than one. Now, we see that the different charges of the $\Psi_i$ might induce a hierarchy among the respective masses.

In the same way, we can write down the Yukawa coupling terms responsible for the Dirac masses, where we assign the additional charges  $q^\ell_i$ to the corresponding lepton doublets $\ell _i$. Again, a simple mass term as in eq.(\ref{eq 5d Yuk int}) is in general not allowed, but under the same assumptions as above we may write down effective Yukawa terms such as
\begin{equation}\label{eq dirac mass u(1)_F}
	- \frac {g_{ij}}{\sqrt{M_5}}
    \left( \frac { \phi}{M_5^{3/2}} 		
    \right)^{q^{\Psi^c}_i+q^\ell_j}
        \overline{\Psi}_i P_L H^T  \left( (\epsilon \ell _j)^T,0,0
        \right)^T
    \delta (y)+ \hc \, ,
	\end{equation}
where the VEV of $\phi $ generates the Dirac masses.


\section{Five-Dimensional Neutrino Mass Models}

Having looked inside mass hierarchies due to an Abelian family symmetry and the principle of SRND,
we can also understand a potential problem of the combination of these two for certain classes of sets of family charges. Let us assume $q^{\Psi^c}_i,q^\ell_j >0$, for the moment. To achieve a mass hierarchy among the SM gauge singlets
(for SRND) one might assign different $\UF$ charges to them, as motivated by eq.(\ref{eq majaorana mass
  u(1)_F}) (or its four-dimensional counterpart). However, in this case the couplings of these singlets to the SM lepton doublets will also be hierarchical, if they are due to terms like eq.(\ref{eq dirac mass u(1)_F}). We can also see that, due to the hierarchy in
the SM gauge singlet sector, the corresponding Majorana mass matrix would already be approximately diagonal. All this combined leads to the fact, that lighter SM gauge singlets would couple less strongly to the SM lepton doublets than the heavier ones. On a more quantitative level a naive calculation with eqs.(\ref{eq majaorana mass
  u(1)_F}), (\ref{eq dirac mass u(1)_F}) and the four-dimensional see-saw formula, 
\begin{equation}
	m_{LL} = m_D M^{-1} m_D^T \, ,
	\end{equation}
shows, that these two hierarchies would tend to annihilate each other in this case, which in return would seem to disfavor SRND. This can be seen even more explicitely in section \ref{first model}, where we will come back to this point, and in section \ref{sec sec tanh mod}, where one can basically apply the same arguments. While the presence of negative charges (for $\Psi ^c$ or $\ell$) under  $\UF$ can already invalidate this argument, we can also see that a deviation from the four-dimensional see-saw formula might also solve this problem.
In sections \ref{first model} and \ref{sec sec tanh mod} we explicitely study models, where this might be the case, while in sections \ref{sec sec tan mod} and \ref{sec first tan mod} we consider further possible models under the influence of the five-dimensional see-saw.

To be specific, we consider several five-dimensional models with an additional continuous symmetry $U(1)_F$, where the SM particles are constrained to a four-dimensional brane while three additional SM gauge singlets can propagate in the bulk of a compactified extra dimension as already considered in previous sections. The models are considered for a variety of different charges of the additional SM gauge singlets and the SM lepton doublets. In all the considered models we assume one or several of the masses of the additional SM gauge singlets to be in a region where the cotangent or hyperbolic cotangent functions of the five-dimensional see-saw mechanism (see eq.(\ref{eq 5d see-saw gen})-(\ref{eq cot see-saw})) might lead to deviations from the corresponding four-dimensional formalism. We also show, for each model, that one can rather easily produce the observed values for the light neutrino masses and mixing angles under the respective assumptions about the Yukawa couplings and the heavy Majorana masses of the SM gauge singlets. While we make somewhat stronger assumptions about the size of some of these masses in the first two models, our corresponding assumptions in the two other models seem less stringent. 

In the first two models we additionally assume $M_S$ to be of negligible size (see eq.(\ref{eq bulk action})). Therefore the dominating mass term will be the vectorlike $M_V$, which violates Lorentz invariance in the additional dimension. While this might be considered as a disadvantage by the reader, this is not the case for the two further models we consider, where scalarlike mass terms for the SM gauge singlets now determine the effective light neutrino mass matrices. Yet, parts of the analysis of the first two models are also valuable for the analysis of the latter ones.

To keep things simple, we assume to work in a basis where the charged lepton mass matrix is already diagonal.

\subsection{First Vectorlike Mass Model}\label{first model}

This model is a first example, how the five-dimensional see-saw mechanism can help to establish SRND in a model with charge assignments, that would tend to prevent this in four dimensions. 

We assign the charges
\begin{equation}
	\begin{array}{rrr}
	\Psi^c_1:r \, ,\quad &\Psi^c_2 :s \, ,\quad &\Psi^c_3:0 \, , \\
	\ell _1:v \, ,\quad  &\ell _2:w \, ,\quad  &\ell_3:w \, ,\hspace{0ex}
	\end{array}
	\end{equation}
under the additional symmetry $\UF$ to the respective particles and additionally
assume $r > s > 0$ and $v>w$. The condition that all terms are
invariant under the symmetry $U(1)_F$ could yield mass terms as in eq.(\ref{eq majaorana mass
  u(1)_F}), in which case the symmetric Majorana mass matrix $M$ for the SM gauge singlets generally has entries of the order of
\begin{equation}
	M \approx M_\Psi \left( \begin{array}{ccc}
		\eps ^{2r} & \eps ^{r+s} & \eps ^r\\
		\cdot & \eps ^{2s} & \eps^s\\
		\cdot & \cdot &1
		\end{array} \right) \, ,
	\end{equation}
where the entries can still differ by order-one pre-factors. Without tuning this matrix has the eigenvalues $M_1 \approx M_\Psi \eps^{2r},M_2\approx M_\Psi \eps^{2s},M_3\approx M_\Psi$ and is diagonalized by a matrix $U$ of the form
\begin{equation}
	U \approx \left( \begin{array}{ccc}
		1 & \eps ^{r-s} & \eps ^r\\
		\eps^{r-s} & 1  & \eps ^s\\
		\eps^r & \eps ^s & 1
		\end{array} \right) \, ,
	\end{equation}	
where again only orders of magnitude are denoted. Under similar assumptions eq.(\ref{eq dirac mass u(1)_F}) can yield a Dirac mass matrix
\begin{equation}
	m_D \approx \frac {v}{\sqrt{\pi R M_5}}\left( \begin{array}{ccc}
		\eps ^{v+r} & \eps ^{v+s} & \eps ^v \\
		\eps ^{w+r} & \eps ^{w+s} & \eps ^w \\
		\eps ^{w+r} & \eps ^{w+s} & \eps ^w 
		\end{array} \right)
	\end{equation}	
for the neutrinos. We also find
\begin{equation}
	m_D U^* \approx m_D \, ,
	\end{equation}
in terms orders of magnitude, which is relevant regarding eq.(\ref{eq cot see-saw}).

To give a first illustration of the possible impact of an extra dimension, we now assume two of the Majorana masses $M_i$ to be approximately equal to an uneven multiple of $1/(2R)$. For reasons of simplicity we additionally choose $M_3$ to be so close to one of these values, that it can be neglected for the remaining analysis of this model.%
\footnote{Due to this suppression the condition $r>s$ could also be inverted, as relabeling would lead to the exact same problem in this case.}
We already point out, that we will also consider less constrained parameter ranges for these values in the other considered models.%

This leads to the following Majorana mass matrix for the light neutrinos (again only denoting orders of magnitude)
\begin{equation}\label{eq first model mLL orders}
	m_{LL} \hspace{0ex}\approx\hspace{0ex} \frac
        {v^2 \eps^{2w}}{2M_5}\hspace{-0ex}\left[ \cot(\pi M_1 R) \eps^{2r}
          \hspace{-0ex}
	\left( \hspace{-1ex}\begin{array}{c}
		\eps ^{v-w} \\ 1 \\ 1 \end{array} \hspace{-1ex}
            \right) \hspace{-1ex} \otimes
	(\eps ^{v-w}, 1 , 1)
	 +\cot(\pi M_2 R)\eps^{2s} \hspace{-0ex}
	 \left( \hspace{-1ex}\begin{array}{c}
		\eps ^{v-w} \\ 1 \\ 1 \end{array}\hspace{-1ex} \right) 	
		\hspace{-1ex}\otimes
	(\eps ^{v-w}, 1 , 1)\right],
	\end{equation}
where we can already see the relatively large 23-sub-matrix, leading to the desired angles for $\theta _{23}$ and $\theta _{13}$. This formula also helps to understand, why SRND seems disfavored in the corresponding four-dimensional model as we mentioned earlier in this section and in the introduction: To see this, all we need to do is to take $1/R$ to infinity, while keeping $M_1$ and $M_2$ fixed. This does not reproduce the four-dimensional case in all details, however, it is sufficient in this context. In this case we can substitute $\cot(\pi M_i R)\rightarrow (\pi M_i R)^{-1}$ in which case both contributions to $M_{LL}$ would tend to be of the same order of magnitude. These considerations would also hold for the contribution of $M_3$.

To understand the parameter influences on the observables better, we rewrite this equation as
\begin{equation}\label{eq mll ag}
	m_{LL} = \eps ^{2(w+r)} \frac {v^2}{2M_5}
	\cot(\pi M_1 R)\underbrace{\left[
	\left( \begin{array}{ccc}
		a^2 \delta ^2 & ab\delta &ac\delta \\
		ab\delta & b^2 &bc \\
		ac\delta & bc &c^2 \end{array} \right) +
	\alpha \left( \begin{array}{ccc}
		e^2 \delta ^2 & ef\delta &eg\delta \\
		ef\delta & f^2 &fg \\
		eg\delta & fg &g^2 \end{array} \right)\right]}_{\equiv \Lambda} \, ,
	\end{equation}
where we stopped to denote only the orders of magnitude of the matrix entries. Therefore we introduced the Latin parameters ($a,b,c,e,f,g$), which should all be typically of order one, while the parameters which are represented by the Greek letters $\delta\equiv \eps^{v-w} $ and $\alpha \equiv \eps ^{-2(r-s)} \cot (\pi M_2R)/\cot (\pi M_1R)$ are assumed to be small and will be considered as perturbation parameters.

The suppression of the contribution of $\Psi_3$ leads to the fact that the lightest neutrino will be massless. Since we can always set the absolute mass scale by adjusting $M_5$, we will from now on only be concerned with the ratio of the two heavier of the light masses and the various mixing angles. Therefore, it is sufficient to consider the matrix $\Lambda $ as defined above.

The eigenvalues of this matrix are
\begin{eqnarray}
	\lambda _1 &=&0 \, , \label{eq lam1 first}\\
	\lambda _2 &=&\alpha \frac{(cf-bg)^2}{b^2+c^2}
		+{\cal O}(\alpha ^m \delta ^n, m+n \ge 2) \, ,\label{eq lam2 first}\\
	\lambda _3 &=& b^2+c^2+\alpha \frac{(bf+cg)^2}{b^2+c^2}
		+{\cal O}(\alpha ^m \delta ^n, m+n \ge 2)\label{eq lam3 first} \, .
	\end{eqnarray}
At this point we can already see, that $\alpha $ gives an approximation for the ratio $m_2/m_3$. However, there is still a lot of freedom for fixing the various order one parameters. We can find further restrictions for them when considering the mixing angles. Therefore we consider the eigenvectors of $\Lambda$. To respective leading order in each entry%
\footnote{In some of the considered models (e.g in this one) we have more than one small parameter. In this case we usually give the corrections for which the sum of the two powers of these two parameters ($\alpha$ and $\delta$) is the lowest. Of course, this ignores the fact that it can in principle happen that there might be a hierarchy among these two parameters, which could make a different ordering of perturbative terms more sensible.} %
 (partially with additional corrections) they are given by
\begin{equation}
	\vec v_1=\left(\begin{array}{c}
		1 \\
		\frac{ce-ag}{bg-cf}\delta\\
		\frac{be-af}{cf-bg}\delta
		\end{array}\right) \, ,
	\vec v_2=\left(\begin{array}{c}
		\frac {b^2e+c^2e-abf-acg}{c^2f-bcg}\delta
		\\
		1\\
		-\frac bc +
		\frac {(bg-cf)(bf+cg)}{c^2(b^2+c^2)}\alpha
		\end{array}\right) \, ,
	\vec v_3=\left(\begin{array}{c}
		\frac ab \delta \\
		1\\
		\frac cb + \frac {(bg-cf)(bf+cg)}{b^2(b^2+c^2)}\alpha
		\end{array}\right) \, .
	\end{equation}
We can now use eqs.(\ref{eq tan23}), (\ref{eq tan13}), and (\ref{eq tan12}) to determine the approximate mixing angles
\begin{eqnarray}
	\tan (\theta _{23}) &\approx& \frac bc
	-\frac {(bg-cf)(bf+cg)}{c^2(b^2+c^2)}\alpha  \, ,\label{eq approx 23}\\
	\tan (\theta _{13}) &\approx& \pm \frac a{\sqrt{b^2+c^2}}\delta \, ,
	\label{eq approx 13}	\\
	\tan (\theta _{12}) &\approx& \sqrt{2}
		\frac c{bg-fc}	\left(
		a \frac {cg+bf}{c^2+b^2}-e \right) \delta \, ,
	\label{eq approx 12}
	\end{eqnarray}
where the sign in eq.(\ref{eq approx 13}) is the same as the one of $b/c$.

\subsubsection{Suitable Parameter Values} \label{sec suit par val}
Here, we want to take a closer look at the restrictions imposed on the various parameters by observation and find guidelines that lead to suitable parameter values. For this task, we restrict ourselves to the constraints set by eq.(\ref{eq obs values}) and assume further possible connections between the parameter bounds to be negligible within this context. 

We first note, that for our approximations that lead to the previous results to be valid, we need $\alpha, \delta
\lesssim 0.5$, while smaller values are preferable. For the same reasons (as well as naturalness) it seems desirable for all other parameters to be of order one.

First let us look at $\alpha $. If the parameters are all of order one and not tuned, we can approximate $\alpha \approx m_2/m_3$ (compare eqs.(\ref{eq lam2 first}) and (\ref{eq lam3 first})). In the case of hierarchical light neutrino masses eq.(\ref{eq obs values}) tells us
$0.15 \lesssim m_2/m_3 \lesssim 0.25$, from which we see that $\alpha $ should be approximately in the same range.

From eq.(\ref{eq obs values}) we also find $0.7 \lesssim |\tan (\theta _{23})| \lesssim 1.4$. Together with eq.(\ref{eq approx 23}) this leads to $b \approx c$. We can always put $c=1$ by redefining the overall mass scale. In this case, we also find $b \approx 1$.

From $|\tan (\theta _{13})| \lesssim 0.2$ and eq.(\ref{eq approx 13}) we find $a \delta \lesssim 0.2$, which does not impose very strong conditions on either of these parameters.

Finally, we have $0.54 \lesssim \tan (\theta _{12}) \lesssim 0.78$ and eq.(\ref{eq approx 12}).
We see that a small $bg-fc$ could make $\tan (\theta _{12})$ sizeable, however this would also decrease $\lambda _2$ and therefore require $\alpha $ to become larger. A sizeable $\tan (\theta _{12})$ as well as $m_2/m_3$ (with small $\alpha $) can yet be achieved, if $bg-fc$ is of order one. Since we have $b \approx c$ from above, as well as $f$ and $g$ of order one, this can be achieved by fulfilling the condition $fg<0$ and $0.5 \lesssim (|f|,|g|) \lesssim 1.5$. Since in the last paragraph we found $a \delta \lesssim 0.2$, we can now approximate 
eq.(\ref{eq approx 12}) further:
\begin{equation}
	\tan (\theta _{12}) \approx -\sqrt{2}
		\frac e{bg-fc} \delta \, .
	\end{equation}
With $b$ and $f$ of order one due to the above argument, this leaves us with the only somewhat stronger condition $0.5 \lesssim e\delta \lesssim 1$.

A suitable set of parameters that agrees with the above conditions is
\begin{equation}\label{eq fst model par val}
	a = 0.5 \, ,\, b = 1 \, ,\, c = 1 \, ,\, e = 3.2 \, ,\, f = 1 \, ,\, g = -0.5 \, ,\, 
	\delta = 0.2 \, ,\, \alpha = 0.3 \, ,
	\end{equation}
yielding the numerical results
\begin{equation}
	\tan (\theta _{23})\approx 1.17 \, ,\quad
	\tan (\theta _{13})\approx 0.12 \, ,\quad
	\tan (\theta _{12})\approx 0.57 \, ,\quad
	m_2/m_3\approx0.21 \, ,
	\end{equation}
which also agree fairly well with our approximative formulae.
	

\subsection{Second Vectorlike Mass Model}\label{sec sec tan mod}

In this section we consider a  model, where an according assignment of $U(1)_F$ charges (combined with eq.(\ref{eq majaorana mass u(1)_F})) leads to a quasi-degeneracy of two of the heavy Majorana masses. This will enable us to fix two of the SM gauge singlet Majorana masses at values close to an uneven multiple of $1/(2R)$ by only adjusting one scale. Therefor, we make the new charge assignments
\begin{equation}
	\begin{array}{rrr}
	\Psi^c_1:r \, ,\quad &\Psi^c_2 :s \, ,\quad &\Psi^c_3:-s \, ,  \\
	\ell _1:v \, ,\quad  &\ell _2:w \, ,\quad  &\ell_3:w \, ,\hspace{2ex}
	\end{array}
	\end{equation}
with $v>w\ge s>0$ and $r>s$.

Again assuming the validity of eq.(\ref{eq majaorana mass u(1)_F}), this leads to the symmetric Majorana mass matrix
\begin{equation}
	M\approx M_\Psi \left( \begin{array}{ccc}
		\eps ^{2r} & \eps ^{r+s} & \eps ^{r-s}\\
		\cdot & \eps ^{2s} & 1\\
		\cdot & \cdot &\eps ^{2s}
		\end{array} \right)
	\end{equation}
for the SM gauge singlets, where we again only denote orders of magnitude for the moment. Without further tuning the eigenvalues of this matrix are typically $M_1\approx M_\Psi\eps ^{2r},M_2\approx M_\Psi,M_3 \approx -M_\Psi$ and the matrix $U$ that diagonalizes $M$ will have entries of order
\begin{equation}\label{eq_2nd_U}
	U \approx \left( \begin{array}{ccc}
		1 & \eps ^{r-s} & \eps ^{r-s}\\
		\eps ^{r-s}& 1  & 1 \\
		\eps^{r+s} & 1  & 1
		\end{array} \right).
	\end{equation}	
In case the Dirac masses are given by eq.(\ref{eq dirac mass u(1)_F}) the Dirac mass matrix $m_D$ is given by
\begin{equation}
	m_D \approx \frac {v}{\sqrt{\pi R M_5}}\left( \begin{array}{ccc}
		\eps ^{v+r} & \eps ^{v+s} & \eps^{v-s} \\
		\eps^{w+r} & \eps^{w+s}  & \eps^{w-s}\\
		\eps^{w+r} & \eps^{w+s}  & \eps^{w-s}
		\end{array} \right) \, ,
	\end{equation}
from which we find
\begin{equation}\label{eq m_DU 2nd}
	m_D U^* \approx \frac {v}{\sqrt{\pi R M_5}}
	\left( \begin{array}{ccc}
		\eps ^{v+r} & \eps^{v-s}& \eps^{v-s} \\
		\eps ^{w+r} & \eps^{w-s}& \eps^{w-s} \\
		\eps ^{w+r} & \eps^{w-s}& \eps^{w-s} \end{array} \right) \, ,
	\end{equation}
again in terms of orders of magnitude for both cases.

Let us for the moment assume that $M_\Psi$ is approximately of order $1/(2R)$ and that the two contributions of $\Psi_2$ and $\Psi_3$ to the five-dimensional see-saw are strongly suppressed. For an arbitrarily strong suppression additional tuning can be necessary (especially for the ratio of the 22- and 33-entries of $M$). However, we also consider the untuned case (in this sense) at a later point. In the additional case that one of the two contributions is suppressed more strongly than the other one, the effective light Majorana mass matrix is approximately
\begin{equation}\label{eq mLL 2nd short}
	m_{LL} \hspace{-0.5ex}\approx\hspace{-0ex} \frac
        {v^2 \eps^{2w}}{2M_5}\hspace{-0.5ex}
        \left[\hspace{-0.5ex}\cot(\pi M_1 R) \eps^{2r}\hspace{-0ex}
	\left( \hspace{-1ex}\begin{array}{c}
		\eps ^{v-w} \\ 1 \\ 1 \end{array} \hspace{-1ex}\right)
            \hspace{-1ex}\otimes  \hspace{-0ex}
	(\eps ^{v-w}, 1 , 1)
	 +\hspace{-0ex} \cot(\pi M_{2} R) \eps^{-2s}\hspace{-0ex}
	 \left( \hspace{-1ex}\begin{array}{c}
		\eps ^{v-w} \\ 1 \\ 1 \end{array} \hspace{-1ex}\right) \hspace{-1ex}\otimes  \hspace{-0ex}
	(\eps ^{v-w}, 1 , 1)\right],
	\end{equation}
where we need to replace $M_2$ by $M_3$ if the contribution of the latter is the more important one (i.e. less suppressed).

Now, if we redefine $\alpha $ by $\alpha \equiv \eps ^{-2(r+s)} \cot (\pi M_2R)/\cot (\pi M_1R)$, we can go back to eq.(\ref{eq mll ag}).
Of course, this also leads to the same analysis for the various parameters and as an example for suitable parameters one can just take the values from section \ref{sec suit par val}.

In the case in which the contributions of $\Psi_2$ and $\Psi_3$ in eq.(\ref{eq cot see-saw}) are suppressed with the same strength, we can write
\begin{equation}\label{eq mll 2nd}
	\begin{array}{l}
	m_{LL} = \frac {v^2}{2M_5}\cot(\pi M_1 R) \eps^{2(w+r)} \vspace{1ex}\\
	\hfill \times \underbrace{\left[
	\left( \begin{array}{ccc}
		a^2 \delta ^2 & ab\delta &ac\delta \\
		ab\delta & b^2 &bc \\
		ac\delta & bc &c^2 \end{array} \right) +
	\alpha \left( \begin{array}{ccc}
		(e^2+h^2) \delta ^2 & (ef+hj)\delta &(eg+hk)\delta \\
		(ef+hj)\delta & f^2+j^2 &fg+jk \\
		(eg+hk)\delta & fg+jk &g^2+k^2 \end{array} \right)\right]}
		_{\equiv \Lambda} \, ,
	\end{array}
	\end{equation}
where $\alpha$ is again redefined by $\alpha \equiv \eps ^{-2(r+s)} \cot (\pi M_2R)/\cot (\pi M_1R)$, $\delta \equiv \eps^{v-w}$ as before, and all the new parameters represented by Latin letters ($a,b,c,e,f,g,h,j,k$) should again be typically of order one. However, to leading order the parameters $h$, $j$, and $k$ cannot be freely chosen. This can be seen from eqs.(\ref{eq_2nd_U})-(\ref{eq m_DU 2nd}), where it becomes obvious that the second and third column in eq.(\ref{eq m_DU 2nd}) are linearly dependent to leading order (a more detailed analysis shows, that they only differ by a minus sign). In fact, for large enough $s$ we can neglect the relative differences of the triples $(e,f,g)$ and $(h,j,k)$. In this case the analysis is again the same as in the previous section after eq.(\ref{eq mll ag}), with the difference that an extra factor of two enters the definition of $\alpha$ in this case compared to the one from eq.(\ref{eq mll 2nd}) ($\rightarrow \alpha \equiv 2 \eps ^{-2(r+s)} \cot (\pi M_2R)/\cot (\pi M_1R)$).%
\footnote{The factor of two is, of course, only exact for $|M_2|=|M_3|$. Depending on the accuracy of this relation $\alpha$ might acquire a different pre-factor}  So, the parameter set from section \ref{sec suit par val} can again be used as an example for a suitable parameter set.

Also in the untuned case (in the sense of the discussion before eq.(\ref{eq mLL 2nd short})), there can be a deviation from the four-dimensional scenario. Let us again assume that either $M_2$ or $M_3$ is equal to $1/(2R)$ up to a degree where it can be neglected. In this case eq.(\ref{eq mLL 2nd short}) is still correct (again a replacement of $M_2$ with $M_3$ can be necessary). Since the difference $|M_2R|-|M_3R|$ is now typically of order $\eps ^{2s}$, we can approximate $\cot (\pi M_2 R) \approx \eps ^{2s}\pi$ for large enough $s$. Also, for large enough $r$, we can approximate $\cot (\pi M_1 R) \approx \eps ^{-2r}/\pi$. In this case, we see that the suppression of the first term with respect to the second one should be approximately $1/\pi^2$. However, a slight variation of $M_1$ and $M_2$ by a respective, approximate factor of $\sqrt{\pi}$ can easily lead to a suppression factor of 0.3. This is the same suppression factor as we used in eq.(\ref{eq fst model par val}). In fact, one can now use all the parameters from eq.(\ref{eq fst model par val}) (with interchanging $a\leftrightarrow e, b\leftrightarrow f,c\leftrightarrow g$ due to the suppression of the first term instead of the second) as an example for a suitable parameter set for this model. 

Let us mention that such a slight variation of $M_2$ and $M_3$ might also lead to a suitable SRND case in four dimensions for this model. However, we see that in the five-dimensional case the hierarchy of the contributions of the SM gauge singlets can be altered with respect to the four-dimensional case, where the second and third generation would typically yield contributions to the effective light neutrino mass matrix that are both larger than the contribution of the first generation. Therefore the fifth dimension can still have an impact on this model.

\subsubsection{Flipped Charge Hierarchy}

Let us also consider the case of an inverted hierarchy of the charges under $U(1)_F$ in the heavy sector ($s>r>0$), while we still assume $v > w \ge s$. Denoting only orders magnitude and making the same assumptions about the mass term structure as before, we typically find
\begin{equation}
	M\approx M_\Psi \left( \begin{array}{ccc}
		\eps ^{2r} & \eps ^{s+r} & \eps ^{s-r}\\
		\cdot & \eps ^{2s} & 1\\
		\cdot & \cdot &\eps ^{2s}
		\end{array} \right)
	\end{equation}
for the entries of the heavy Majorana mass matrix. The eigenvalues of this matrix are again $M_1 \approx \eps ^{2r},M_2 \approx M_\Psi,M_3 \approx -M_\Psi$, while the matrix $U$ by which it is diagonalized now generally has entries of order
\begin{equation}
	U \approx \left( \begin{array}{ccc}
		1 & \eps ^{s-r} & \eps ^{s-r}\\
		\eps ^{s-r}& 1  & 1 \\
		\eps^{s+r} & 1  & 1
		\end{array} \right) \, .
	\end{equation}
Once more assuming the validity of eq.(\ref{eq dirac mass u(1)_F}), the Dirac mass matrix $m_D$ now has typical entries of order
\begin{equation}
	m_D \approx \frac {v}{\sqrt{\pi R M_5}}\left( \begin{array}{ccc}
		\eps ^{v+r} & \eps ^{v+s} & \eps^{v-s} \\
		\eps^{w+r} & \eps^{w+s}  & \eps^{w-s}\\
		\eps^{w+r} & \eps^{w+s}  & \eps^{w-s}
		\end{array} \right)
	\end{equation}
and multiplying it with $U$ from the right-hand side generally yields
\begin{equation}
	m_D U^*\approx \frac {v}{\sqrt{\pi R M_5}}\left( \begin{array}{ccc}
		\eps ^{v+r} & \eps ^{v-s} & \eps^{v-s} \\
		\eps^{w+r} & \eps^{w-s}  & \eps^{w-s}\\
		\eps^{w+r} & \eps^{w-s}  & \eps^{w-s}
		\end{array} \right) \, .
	\end{equation}
Comparing this with eq.(\ref{eq m_DU 2nd}), we see that we are again lead to equations (\ref{eq mLL 2nd short}) and/or (\ref{eq mll 2nd}) and the analysis that follows them remains unchanged. This is even true for the untuned case (again in the sense of the discussion before eq.(\ref{eq mLL 2nd short})), since an analysis shows that $|M_2R|-|M_3R|$ will again be of order $\eps^{2s}$ under the previous assumption $M_{2/3}\approx 1/(2R)$.

\subsection{First Scalarlike Mass Model}\label{sec first tan mod}

Next, we consider a model that makes use of a scalarlike extra-dimensional mass term and therefore of the hyperbolic cotangent function
of eq.(\ref{eq coth see-saw}). While SRND would also be possible in four dimensions for this model, the five-dimensional see-saw can again alter the structure of the effective mass matrix of the light neutrinos. 

We now assume our Majorana mass term for the SM gauge singlets to be due to the scalar mass term $M_S$ in eq.(\ref{eq bulk action}) and use the charge assignments
\begin{equation}
	\begin{array}{rrr}
	N_1:r' \, ,\quad &N_2 :0 \, ,\quad &N_3:0 \, , \\
	\ell _1:r' \, ,\quad &\ell _2:-r' \, ,\hspace{-2ex}\quad  &\ell_3:-r' \, ,\hspace{-2ex}
	\end{array}
	\end{equation}
with $r'\not=0$. Under our previous assumptions, this leads to a SM gauge singlet Majorana mass matrix $M$ with typical entries of the order
\begin{equation}
	M\approx M_\Psi \left( \begin{array}{ccc}
		\eps ^{2r} & \eps ^{r} & \eps ^{r}\\
		\cdot & 1 & 1\\
		\cdot & \cdot &1
		\end{array} \right) \, ,
	\end{equation}
with $r\equiv |r'|$.

Typically, the corresponding eigenvalues are now of the respective orders $M_1\approx M_\Psi \eps^{2r},M_2\approx M_\Psi,M_3 \approx M_\Psi$, while the matrix $U$ that diagonalizes $M$ generally has entries of order
\begin{equation}
	U \approx \left( \begin{array}{ccc}
		1 & \eps ^{r} & \eps ^{r}\\
		\eps ^{r}& 1  & 1 \\
		\eps^{r} & 1  & 1
		\end{array} \right) \, .
	\end{equation}	
Also, under the same assumptions as in the earlier sections, the Dirac mass matrix $m_D$ now takes the form
\begin{equation}
	m_D \approx \frac {v}{\sqrt{\pi R M_5}}\left( \begin{array}{ccc}
		\eps ^{2r} & \eps ^{r} & \eps^{r} \\
		1 & \eps^{r}  & \eps^{r}\\
		1 & \eps^{r}  & \eps^{r}
		\end{array} \right) \, ,
	\end{equation}
which leads to $m_D U^* \approx m_D$ with respect to orders of $\eps$ for the respective entries. 

If we now take $M_\Psi\gg 1/R$ we can set the hyperbolic cotangent functions in eq.(\ref{eq coth see-saw}) equal to one, which in return leads to the effective mass matrix 
\begin{equation}\label{eq mLL 3rd short}
	m_{LL} \approx \frac {v^2}{2M_5} \left[
	\left( \begin{array}{c}
		\eps ^{2r} \\ 1 \\ 1 \end{array} \right)\otimes
	(\eps ^{2r}, 1 , 1)
	 +  \left( \begin{array}{c}
		\eps ^{r} \\ \eps ^{r} \\ \eps ^{r} \end{array}
            \right) \otimes
	(\eps ^{r}, \eps ^{r} ,\eps ^{r})
	+  \left( \begin{array}{c}
		\eps ^{r} \\ \eps ^{r} \\ \eps ^{r} \end{array}
            \right) \otimes
	(\eps ^{r}, \eps ^{r} ,\eps ^{r}) \right]
	\end{equation}
for the light neutrinos. Similarly to the previous models, we can write the matrix as
\begin{equation}\label{eq lambda def 1st tanh mod}
	m_{LL} = \frac {v^2}{2M_5}
	\hfill \times \underbrace{\left[
	\left( \begin{array}{ccc}
		a^2 \delta ^4 & ab \delta ^2&ac\delta ^2 \\
		ab\delta ^2 & b^2 &bc \\
		ac\delta ^2 & bc &c^2 \end{array} \right) +
	\left( \begin{array}{ccc}
		(e^2+h^2) \delta ^2 & (ef+hj)\delta ^2 &(eg+hk)\delta ^2 \\
		(ef+hj)\delta ^2 & (f^2+j^2)\delta ^2 &(fg+jk)\delta ^2 \\
		(eg+hk)\delta ^2 & (fg+jk)\delta ^2 &(g^2+k^2)\delta ^2
		 \end{array} \right)\right]}
		_{\equiv \Lambda} \, ,
	\end{equation}
with $\delta \equiv \eps^r$. Approximative formulae for eigenvalues and -vectors of this matrix, which can help finding suitable parameters that reproduce the observed values for the light neutrino masses and mixing angles, are rather complex due to the large number of parameters. Therefore we only give an example for a suitable parameter set that reproduces the desired neutrino data at this point, while we present approximative formulae for eigenvalues and -vectors of $\Lambda$ in the appendix.

The parameters
\begin{equation*}
	\begin{array}{llllll}
	a = 0.5 \, ,& b = 1 \, ,& c = 1 \, ,&
	e = 0.6 \, ,& f = 1 \, ,& g = -0.6 \, , \\
	h = -0.5 \, ,& j = 0.9 \, ,& k = 1 \, ,&
	\delta = 0.6 \, ,
	\end{array}
	\end{equation*}
yield
\begin{equation*}
	\tan (\theta _{23})\approx 1.08 \, ,\quad
	\tan (\theta _{13})\approx 0.03 \, ,\quad
	\tan (\theta _{12})\approx 0.67 \, ,\quad
	m_2/m_3\approx0.23 \, .
	\end{equation*}	
We see that all the parameters are approximately of the order of one. The relatively big $\delta$ should be no reason for worry in terms of small perturbations, since only its square and higher orders of it appear in the important equations.


\subsection{Second Scalarlike Mass Model}\label{sec sec tanh mod}
Finally, we take another look at the first model presented in this paper in section \ref{first model}. However, we now consider the case, where the Majorana mass term is due to a scalar mass $M_S$. 

This model has several nice features. As in the case of a vectorlike mass-term in section \ref{first model} one can obtain SRND through the five-dimensional see-saw. Additionally, we do not need to assume that two of the Majorana masses are approximately equal to an uneven multiple of $1/(2R)$ in this model, as we will see. Due to these features and the five-dimensional Lorentz invariance, this could be considered as the most natural of our considered models.

In case of a scalarlike mass term and under similar assumptions as before eq.(\ref{eq first model mLL orders}) becomes
\begin{eqnarray}\label{eq fourth model mLL orders}
	m_{LL} \approx \frac {v^2}{2M_5}\Bigg[ \coth(\pi M_1 R) \eps^{2(w+r)}
	\left( \hspace{-1ex}\begin{array}{c}
		\eps ^{v-w} \\ 1 \\ 1 \end{array} \hspace{-1ex}
            \right) \otimes
	(\eps ^{v-w}, 1, 1) \nonumber\hspace{4ex}& \\
	 +\hspace{-0.5ex}\coth(\pi M_2 R)\eps^{2(w+s)}\hspace{-1ex}
	 \left( \hspace{-1ex}\begin{array}{c}
		\eps ^{v-w} \\ 1 \\ 1 \end{array}\hspace{-1ex} \right)
           \hspace{-1ex} \otimes
	(\eps ^{v-w}, 1 , 1)
	+\hspace{-0.5ex}\coth(\pi  M_3 R)\eps^{2w}&\hspace{-1.5ex}\hspace{-0.5ex}
	 \left( \hspace{-1ex}\begin{array}{c}
		\eps ^{v-w} \\ 1 \\ 1 \end{array}\hspace{-1ex} \right)
            \hspace{-1ex} \otimes
	(\eps ^{v-w}, 1 , 1)
	\Bigg] \, ,
	\end{eqnarray}
where we point out that the relations $M_1\approx \eps^{2r}M_3$ and $M_2\approx \eps^{2s}M_3$ from the corresponding cotangent model generally still hold. Now, we can no longer simply suppress one of the three terms by assuming that the corresponding $M_i$ is very close to an uneven multiple of $1/(2R)$. However, new effects can still arise (compared to the four-dimensional case), if the mass scale $M_\Psi$ comes within the range of $1/R$. 

We can  rewrite eq.(\ref{eq fourth model mLL orders}) as
\begin{eqnarray}
	m_{LL} \approx \frac {v^2\eps^{2w}\coth(\pi M_3 R)}{2M_5}
	\Bigg[ \beta
	 \hspace{-0.5ex}\left( \hspace{-1ex}\begin{array}{c}
		\eps ^{v-w} \\ 1 \\ 1 \end{array} \hspace{-1ex}
            \right)&\hspace{-35ex}\otimes \hspace{1ex}
	(\eps ^{v-w}, 1 , 1)\nonumber \\
	 +\alpha 
	 &\hspace{-2ex}\left( \hspace{-1ex}\begin{array}{c}
		\eps ^{v-w} \\ 1 \\ 1 \end{array}\hspace{-1ex}
            \right)\hspace{-0.5ex} \otimes
	(\eps ^{v-w}, 1 , 1) 
	+
	\hspace{-0.5ex} \left( \hspace{-1ex}\begin{array}{c}
		\eps ^{v-w} \\ 1 \\ 1 \end{array}\hspace{-1ex} \right)\hspace{-0.5ex}\otimes
	(\eps ^{v-w}, 1 , 1)
	\Bigg]  , 
	\end{eqnarray}
where the parameters represented by the Greek letters $\beta \equiv \coth(\pi M_1 R) \eps^{2r}/\coth(\pi M_3 R)$ and $\alpha \equiv \coth(\pi M_2 R) \eps^{2s}/\coth(\pi M_3 R)$ are once again considered as small perturbations. 

From here on, we consider two different cases. First we consider the
case $M_iR \gg 1$ for all $i$. Since $r>s$, we see that, typically, the contribution
of the lightest of the heavy neutrinos to the effective mass matrix of
the light neutrinos is now suppressed the strongest. For a large enough $r$, we can therefore always ignore the contribution of $\Psi_1$, in which case the problem of eigenvalues, -vectors and mixing angles resembles the one in eq.(\ref{eq mll ag}), where of course the parameters $(a,b,c,e,f,g)$ now need to be assigned to different generations, respectively, and we also have a different overall pre-factor for the effective mass matrix. However, an important difference of the present analysis is the fact, that $\alpha$ cannot be freely chosen, since it is now equal to $\eps ^{2s}$. Yet, it is still not hard to find a suitable parameter set: All we now need to do is to fix $\eps$, $s$, and $v-w$ in a way, such that the parameters $\alpha$ and $\delta$ from eq.(\ref{eq fst model par val}) are approximately reproduced. If the result is close enough to the original values all the other parameters from this equation can remain the same or only need to change very little. As an example we found $\eps = 0.58$, $s=1$, $v-w=3$, which yield $\alpha \approx 0.34$, $\delta \approx 0.20$ and
\begin{equation}
	\tan (\theta _{23})\approx 1.20 \, ,\quad
	\tan (\theta _{13})\approx 0.12 \, ,\quad
	\tan (\theta _{12})\approx 0.55 \, ,\quad
	m_2/m_3\approx0.23 \, ,
	\end{equation}
if we do not change the other parameters from eq.(\ref{eq fst model par val}).

The other case we want to look at, is the case where $M_3R\lesssim 1$. For large enough $r$ and $s$ the masses $M_1$ and $M_2$ can be much smaller than $M_3$, such that $\coth(\pi M_iR)$ can be replaced by $1/(\pi M_iR)$ for both of the particles. In this case we find respective cancellations of the factors of $\eps ^{2r}$ and $\eps ^{2s}$ in eq.(\ref{eq fourth model mLL orders}) for the first and second contribution. If this was the case for all three generations, we would again have to face three contributions that are typically of the same order of magnitude as in the corresponding four-dimensional model. However, in our case this only leads to the fact that the first two contributions should typically be suppressed with the same strength with respect to the contribution of the heaviest SM gauge singlet. (For $M_3R\lesssim 1$ the cotangent function yields a value, which is somewhat bigger than one). This again leads to the previously considered eq.(\ref{eq mll 2nd}), where of course the Latin parameters ($a,b,c,e,f,g,h,j,k$) now have to be assigned to different respective generations compared to the former case and can now be chosen freely. Additionally, we find a different overall pre-factor for $\Lambda$ and  $\alpha$ is now defined by $\alpha \equiv \tanh(\pi M_3 R)/\pi$. Approximative formulae for this case can be found in appendix \ref{appendix second vec}. 

An example for a suitable parameter set is
\begin{equation*}
	\begin{array}{llllll}
	a = 0.5 \, ,& b = 1 \, ,& c = 1 \, ,&
	e = 3 \, ,& f = 1 \, ,& g = -0.5 \, , \\
	h = 2.8 \, ,& j = 0.9 \, ,& k = -0.4 \, ,&
	\alpha = 0.2 \, ,& \delta = 0.2 \, ,
	\end{array}
	\end{equation*}
which yields
\begin{equation*}
	\begin{array}{llll}
	\tan (\theta _{23})\approx 1.22 \, ,&
	\tan (\theta _{13})\approx 0.13 \, ,&
	\tan (\theta _{12})\approx 0.55 \, ,&
	m_2/m_3\approx0.24 \, .
	\end{array}
	\end{equation*}	

\section{Conclusions}

In this paper we have treated several models that all illustrated, how extra dimensions might have a considerable impact on models with a continuous $U(1)_F$ family symmetry introduced to explain the structure of the effective Majorana mass matrix of the light neutrinos. In all models the five-dimensional see-saw was used to establish SRND scenarios. In two of the models the five-dimensional see-saw may lead to SRND, where the corresponding four-dimensional models would naively not lead to one dominating contribution to the effective Majorana mass matrix of the light neutrinos by a single right-handed neutrino.

In our particular models the SM particles were confined to a brane, while three additional gauge singlets were introduced, that could propagate in the bulk. The models differed in the origin of the heavy Majorana mass term (vectorlike and scalarlike) and in the various charges assigned to the SM lepton doublets and the additional SM gauge singlets under the new symmetry. 

For the models with a vectorlike Majorana mass term for the SM gauge singlets we made assumptions that some of the Majorana masses are similar to an uneven multiple of $1/(2R)$. For these assumptions some further motivation would be desirable. However, also for other values of $M_i$, which we did not consider, there might be a significant deviation from the inversely linear behavior due to the cotangent function.

For the models with a scalarlike Majorana mass term for the SM gauge singlets, we considered several cases for which the assumptions about the $M_i$ seem more motivated, since it does not seem like a very strong assumption that all Majorana masses are at a scale that is much higher than the scale set by $1/R$, for instance. However, we also considered parameter values where this is not the case.

All models were treated on a quantitative level and approximative formulae that can be helpful for the determination of neutrino masses and mixings were presented. For each of the models, we also presented parameter sets that yield the observed values for neutrino mass ratios and mixing angles under the respective assumptions and that fulfill the naturalness condition that all of the coupling constants are roughly of the same order of magnitude.  

Since all these  models help to demonstrate the possible huge importance of extra dimensions for phenomenology and model building further investigation on this topic is desirable.


\section*{Acknowledgments}
We would like to thank M. Lindner, E. Akhmedov, C. Hagedorn, A. Merle, N. Okada, T. Ota, W. Rodejohann, M. Rolinec, M. Schmidt, T. Yamashita, and K. Yoshioka for helpful discussions. M.T.E. wishes to thank the {}``Graduiertenkolleg
1054'' of the {}``Deutsche Forschungsgemeinschaft'' for financial
support. N.H. is supported by the Alexander von Humboldt Foundation. 

\newpage
\begin{appendix}
\section{Mixing Angles}
Here, we re-derive some formulae for the mixing angles that might be convenient in the context of this paper. The argument is along Ref. \cite{King:1999mb}.

The matrix $m_{LL}$ is symmetric and can therefore be diagonalized by a unitary matrix $V$:
\begin{equation}\label{eq mLL diag}
	m^D_{LL}=V m_{LL} V^T  \, ,
	\end{equation}
where $m^D_{LL}\equiv \mathrm{diag}(m_1,m_2,m_3)$ with $m_1<m_2<m_3$.

Since this work is not concerned with CP phases, we assume $m_{LL}$ to be real. Hence, $V$ is real as well and can be expressed through three consecutive rotations around the three axes
\begin{equation}
	V^T=R_{23}R_{13}R_{12} \, ,
	\end{equation}
with
\begin{equation}\label{eq eigenvectors 1st model}
	\begin{array}{c}
	R_{23}\equiv \left( \begin{array}{ccc}
		1&0&0\\
		0& \cos (\theta _{23})&\sin (\theta _{23})\\
		0&-\sin (\theta _{23})& \cos (\theta _{23})
		\end{array}\right) \, , \quad \vspace{1ex}
	R_{13}\equiv \left( \begin{array}{ccc}
		\cos (\theta _{13})&0&\sin (\theta _{13})\\
		0&1&0\\
		-\sin (\theta _{13})&0& \cos (\theta _{13})
		\end{array}\right) \, , \vspace{1ex}\\
	R_{12}\equiv \left( \begin{array}{ccc}
		\cos (\theta _{12})&\sin (\theta _{12})&0\\
		-\sin (\theta _{12})& \cos (\theta _{12})&0\\
		0&0&1
		\end{array}\right) \, , \hfill
	\end{array}
	\end{equation}
and the various $\theta _{ij}$ are fixed by eq.(\ref{eq mLL diag}).

On the other hand, $V$ can also be expressed via the three eigenvectors of $m_{LL}$
\begin{equation}
	V^T=(\vec v_1,\vec v_2,\vec v_3) \, ,
	\end{equation}
where each $\vec v_i$ is the normalized eigenvector corresponding to the respective eigenvalue $m_i$.

This however yields
\begin{equation}
	R_{12}^TR_{13}^TR_{23}^T(\vec v_1,\vec v_2,\vec v_3)=
	\mathrm{diag}(1,1,1)
	\end{equation}
and therefore
\begin{equation}
	R_{13}^TR_{23}^T \vec v_3 =
	\left( \begin{array}{c} 0\\0\\1 \end{array}\right)
	\end{equation}
and
\begin{equation}\label{eq rot v2}
	R_{12}^TR_{13}^TR_{23}^T \vec v_2 =
	\left( \begin{array}{c} 0\\1\\0 \end{array}\right) \, .
	\end{equation}
Now we can consecutively determine the various $\theta _{ij}$.

First we see that
\begin{equation}
	R_{23}^T \vec v_3 
	= \left( \begin{array}{c} v_{3,x}\\0\\
	\pm \sqrt{v_{3,y}^2+ v_{3,z}^2} \end{array}\right) \, ,
	\end{equation}
where the sign in front of the expression is the same as the one of $v _{3,z}$.%
This fixes $\theta _{23}$ by the condition
\begin{equation}\label{eq tan23}
	\tan (\theta _{23}) = \frac {v_{3,y}}{v_{3,z}} \, .
	\end{equation}
Next we have
\begin{equation}
	R_{13}^T \left( \begin{array}{c} v_{3,x}\\0\\
	\pm \sqrt{v_{3,y}^2+ v_{3,z}^2} \end{array}\right)
	= \left( \begin{array}{c} 0\\0\\ 
	\pm \sqrt{v_{3,x}^2+v_{3,y}^2+ v_{3,z}^2} \end{array}\right) \, ,
	\end{equation}
which leads to
\begin{equation}\label{eq tan13}
	\tan (\theta _{13}) = \pm \frac {v_{3,x}}{\sqrt{v_{3,y}^2+ v_{3,z}^2}} \, ,
	\end{equation}
where the sign in front of the expression is again the same as the one of $v _{3,z}$. Of course, a negative sign of $\theta _{13}$ could always be absorbed by a CP-phase $\delta$ in a more general consideration.

From the relation
\begin{equation}
	R_{13}^T R_{23}^T \vec v_2 =
	\left( \begin{array}{c}
	c_{13} v_{2,x}-	s_{13} (c_{23} v_{2,z} + s_{23} v_{2,y})\\
	c_{23} v_{2,y}-s_{23} v_{2,z}\\
	c_{13} (c_{23} v_{2,z}+s_{23} v_{2,y})+s_{13} v_{2,x}
	\end{array} \right) \, ,
	\end{equation}
with $c_{ij}\equiv \cos(\theta _{ij})$ and $s_{ij}\equiv \sin(\theta _{ij})$, and eq.(\ref{eq rot v2}) we can now determine $\theta _{12}$ with the formula
\begin{equation}\label{eq tan12}
	\tan (\theta _{12}) = \frac {\cos (\theta _{13}) v_{2,x}
		-\sin (\theta _{13})[\sin (\theta _{23}) v_{2,y}
		+\cos(\theta _{23}) v_{2,z}]}
		{\cos (\theta _{23}) v_{2,y}-\sin(\theta _{23}) v_{2,z}} \, .
	\end{equation}
If we approximate $c_{13} \approx 1$,$s_{13} \approx 0$ and $c_{23}\approx s_{23} \approx 1/\sqrt{2}$, this leads to
\begin{equation}\label{eq tan12 approx1}
	\tan (\theta _{12}) \approx \sqrt{2} \frac {v_{2,x}}
		{v_{2,y}-\tan(\theta _{23}) v_{2,z}} \, .
	\end{equation}

\section{First Scalarlike Mass Model: \newline Eigenvalues and Mixing Angles}
\label{appendix first scalar}

We give some approximative formulae for the first scalarlike mass model from section \ref{sec first tan mod}, that can help finding suitable parameter values. 

For the eigenvalues of $\Lambda $ as defined in eq.(\ref{eq lambda def 1st tanh mod}) we find the approximative formulae
\begin{eqnarray}
	\lambda _1 &\approx&  
		\frac {Y-
		\sqrt{Y^2-4XZ}}{2X}
		\delta ^2 \, ,\\
	\lambda _2 &\approx&\frac {Y+
		\sqrt{Y^2-4XZ}}{2X}
		\delta ^2 \, ,\\
	\lambda _3 &\approx& X + \frac {b^2(f^2+j^2)+2bc(fg+jk)+c^2(g^2+k^2)}{X} 		\delta ^2 \, ,
	\end{eqnarray}	
where we only give the leading contributions for each value with
\begin{eqnarray}
	X &\equiv&  b^2+c^2 \, ,\\
	Y &\equiv& b^2e^2+c^2e^2+c^2f^2-2bcfg+b^2g^2+b^2h^2+c^2h^2+c^2j^2
		-2bcjk+b^2k^2 \, ,\\
	Z &\equiv& (cfh-bgh-cej+bek)^2 \, .
	\end{eqnarray}
With these values we find the approximate, corresponding eigenvectors
\begin{equation}
\begin{array}{c}
	\vec v_1\approx\left(\begin{array}{c}
		1 \\
		\frac{c(e^2+h^2+
		(-Y+\sqrt{Y^2-4XZ})/(2X))}{-c(ef+hj)+b(eg+hk)}\\
		b\frac{-c^2(e^2-f^2+h^2-j^2)-2bc(fg+jk)-b^2(e^2-g^2+h^2-k^2)
		-\sqrt{Y^2-4XZ}}{2X(-c(ef+hj)+b(eg+hk))}
		\end{array}\right) \, ,\vspace{2ex}
		\\
	\vec v_2\approx\left(\begin{array}{c}
		\frac{-c(ef+hj)+b(eg+hk)}{c(e^2+h^2-
		(Y+\sqrt{Y^2-4XZ})/(2X))}
		\\
		1\\
		-\frac bc 
		\end{array}\right) \, ,\vspace{2ex}
		\\
	\vec v_3\approx\left(\begin{array}{c}
		\frac{a b^2 + a c^2 + b e f + c e g + b h j + c h k}
			{b^3 + b c^2} \delta ^2\\
		1\\
		\frac cb
		+\frac{b^2(f g + j k) - c^2(f g + j k)+ bc(-f^2+g^2-j^2+k^2)} 					{b^2(b^2 +c^2)}\delta^2
		\end{array}\right) \, .
		\end{array}
	\end{equation}	
One can now use eqs.(\ref{eq tan23}), (\ref{eq tan13}), and (\ref{eq tan12 approx1}) to find approximations for the mixing angles. Here, we only want to point out, that the order one entry of $v_{2,x}$ (if its corresponding eigenvalue is larger than that of $\vec v_1$) makes it in general easier to achieve a sizeable $\theta _{12}$, while the orders of magnitude for the various entries of $\vec v_3$ naturally lead to a large $\theta _{23}$ and a small $\theta _{13}$.

\section{Second Scalarlike Mass Model: \\Eigenvalues and Mixing Angles}
\label{appendix second vec}

For completeness we also give approximate formulae for the eigenvalues of $\Lambda $ as it was defined in eq.(\ref{eq mll 2nd}) and used in section \ref{sec sec tanh mod}, its eigenvectors, and the resulting mixing angles. These formulae can help finding suitable parameter values for the corresponding model.

To respective leading order we find the following approximative eigenvalues for $\Lambda $
\begin{eqnarray}
	\lambda _1 &=& \frac{(cfh-bgh-cej+agj+bek-afk)^2}
		{c^2f^2-2bcfg+b^2g^2+c^2j^2-2bcjk+b^2k^2}
		\alpha \delta ^2 \, ,\\
	\lambda _2 &=&\frac{c^2f^2-2bcfg+b^2g^2+c^2j^2-2bcjk+b^2k^2}
		{b^2+c^2}\alpha \, ,\\
	\lambda _3 &=& b^2+c^2
	 +\frac{b^2f^2+2bcfg+c^2g^2+b^2j^2+2bcjk+c^2k^2}
	 	{b^2+c^2} \alpha \, ,
	\end{eqnarray}
where we also gave the order $\alpha $ contribution for the largest eigenvalue.

For the corresponding eigenvectors we find
\begin{equation}
\begin{array}{c}
	\vec v_1=\left(\begin{array}{c}
		1 \\
		\frac{-c^2(ef+hj)+c(beg+afg+bhk+ajk)-ab(g^2+k^2)}
		{c^2(f^2+j^2)-2bc(fg+jk)+b^2(g^2+k^2)}\delta \vspace{1ex}\\
		\frac{-ac(f^2+j^2)-b^2(eg+hk)+b(cef+afg+chj+ajk)}
		{c^2(f^2+j^2)-2bc(fg+jk)+b^2(g^2+k^2)} \vspace{1ex}\delta
		\end{array}\right) \, ,\vspace{2ex}
		\\
	\vec v_2=\left(\begin{array}{c}
		\frac {-b^3(eg+hk)+b^2(cef+afg+chj+ajk)
		+c^2(c(ef+hj)-a(fg+jk))
		-bc(c(eg+hk)+a(f^2-g^2+j^2-k^2))}
		{c(c^2(f^2+j^2)-2bc(fg+jk)+b^2(g^2+k^2))}\delta
		\\
		1\\
		-\frac bc 
		+\frac{b^2(fg+jk)-c^2(fg+jk)+bc(-f^2+g^2-j^2+k^2)}
		{c^2(b^2+c^2)}\alpha
		\end{array}\right) \, ,\vspace{2ex}
		\\
	\vec v_3=\left(\begin{array}{c}
		\frac ab \delta \\
		1\\
		\frac cb
		+ \frac{b^2(fg+jk)-c^2(fg+jk)+bc(-f^2+g^2-j^2+k^2)}
		{b^2(b^2+c^2)}\alpha
		\end{array}\right) \, ,
		\end{array}
	\end{equation}	
where we again only give the respective leading order contributions for each entry and the order $\alpha $ corrections to the order one entries $v_{3,z}$ and $v_{2,z}$.

With the same analysis as used before, we therefore find the approximate mixing angles
\begin{eqnarray}
	\tan (\theta _{23}) &\approx& \frac bc
		- \frac{b^2(fg+jk)-c^2(fg+jk)+bc(-f^2+g^2-j^2+k^2)}
		{c^2(b^2+c^2)}\alpha \, , \\
	\tan (\theta _{13}) &\approx& \pm \frac a{\sqrt{b^2+c^2}}\delta \, ,\\
	\tan (\theta _{12}) &\approx& 
		\frac {\sqrt{2}}{1+b^2/c^2}\, \delta \,
		\Big( -b^3(eg+hk)+b^2(cef+afg+chj+ajk) \nonumber \\
		&&	+c^2(c(ef+hj)-a(fg+jk))
		-bc(c(eg+hk)+a(f^2-g^2+j^2-k^2)) \Big) \nonumber \\
		&&\Big/ \Big( c(c^2(f^2+j^2)-2bc(fg+jk)+b^2(g^2+k^2)) \Big) \, .
	\end{eqnarray}

\end{appendix}


\clearpage
\newpage
\bibliographystyle{apsrev}
\bibliography{references}

\begin{mcbibliography}{10}
\expandafter\ifx\csname bibnamefont\endcsname\relax
  \def\bibnamefont#1{#1}\fi
\expandafter\ifx\csname bibfnamefont\endcsname\relax
  \def\bibfnamefont#1{#1}\fi
\expandafter\ifx\csname url\endcsname\relax
  \def\url#1{\texttt{#1}}\fi
\expandafter\ifx\csname urlprefix\endcsname\relax\def\urlprefix{URL }\fi
\providecommand{\bibinfo}[2]{#2}
\providecommand{\eprint}[2][]{\url{#2}}

\bibitem{Maltoni:2004ei}
\bibinfo{author}{\bibfnamefont{M.}~\bibnamefont{Maltoni}},
  \bibinfo{author}{\bibfnamefont{T.}~\bibnamefont{Schwetz}},
  \bibinfo{author}{\bibfnamefont{M.~A.} \bibnamefont{Tortola}},
  \bibnamefont{and} \bibinfo{author}{\bibfnamefont{J.~W.~F.}
  \bibnamefont{Valle}}, \bibinfo{journal}{New J. Phys.}
  \textbf{\bibinfo{volume}{6}}, \bibinfo{pages}{122} (\bibinfo{year}{2004}),
  \eprint{hep-ph/0405172}\relax
\relax
\bibitem{Fogli:2003th}
\bibinfo{author}{\bibfnamefont{G.~L.} \bibnamefont{Fogli}},
  \bibinfo{author}{\bibfnamefont{E.}~\bibnamefont{Lisi}},
  \bibinfo{author}{\bibfnamefont{A.}~\bibnamefont{Marrone}}, \bibnamefont{and}
  \bibinfo{author}{\bibfnamefont{D.}~\bibnamefont{Montanino}},
  \bibinfo{journal}{Phys. Rev.} \textbf{\bibinfo{volume}{D67}},
  \bibinfo{pages}{093006} (\bibinfo{year}{2003}), \eprint{hep-ph/0303064}\relax
\relax
\bibitem{Bahcall:2004ut}
\bibinfo{author}{\bibfnamefont{J.~N.} \bibnamefont{Bahcall}},
  \bibinfo{author}{\bibfnamefont{M.~C.} \bibnamefont{Gonzalez-Garcia}},
  \bibnamefont{and} \bibinfo{author}{\bibfnamefont{C.}~\bibnamefont{Pe\~na
  Garay}}, \bibinfo{journal}{JHEP} \textbf{\bibinfo{volume}{08}},
  \bibinfo{pages}{016} (\bibinfo{year}{2004}), \eprint{hep-ph/0406294}\relax
\relax
\bibitem{Bandyopadhyay:2004da}
\bibinfo{author}{\bibfnamefont{A.}~\bibnamefont{Bandyopadhyay}},
  \bibinfo{author}{\bibfnamefont{S.}~\bibnamefont{Choubey}},
  \bibinfo{author}{\bibfnamefont{S.}~\bibnamefont{Goswami}},
  \bibinfo{author}{\bibfnamefont{S.~T.} \bibnamefont{Petcov}},
  \bibnamefont{and} \bibinfo{author}{\bibfnamefont{D.~P.} \bibnamefont{Roy}},
  \bibinfo{journal}{Phys. Lett.} \textbf{\bibinfo{volume}{B608}},
  \bibinfo{pages}{115} (\bibinfo{year}{2005}), \eprint{hep-ph/0406328}\relax
\relax
\bibitem{Minkowski:1977sc}
\bibinfo{author}{\bibfnamefont{P.}~\bibnamefont{Minkowski}},
  \bibinfo{journal}{Phys. Lett.} \textbf{\bibinfo{volume}{B67}},
  \bibinfo{pages}{421} (\bibinfo{year}{1977})\relax
\relax
\bibitem{Yanagida:1979as}
\bibinfo{author}{\bibfnamefont{T.}~\bibnamefont{Yanagida}} \bibinfo{note}{{} in
  Proceedings of the Workshop on the Baryon Number of the Universe and Unified
  Theories, Tsukuba, Japan, 13-14 Feb 1979, edited by O.Sawada and A. Sugamoto,
  Report KEK-79-18 (1979)}\relax
\relax
\bibitem{Gell-Mann:1980vs}
\bibinfo{author}{\bibfnamefont{M.}~\bibnamefont{Gell-Mann}},
  \bibinfo{author}{\bibfnamefont{P.}~\bibnamefont{Ramond}}, \bibnamefont{and}
  \bibinfo{author}{\bibfnamefont{R.}~\bibnamefont{Slansky}} \bibinfo{note}{{}in
  {\sl Supergravity}, edited by D. Z. Freedman and P. van Nieuwenhuizen
  (North-Holland, Amsterdam, 1979)}\relax
\relax
\bibitem{Arkani-Hamed:1998vp}
\bibinfo{author}{\bibfnamefont{N.}~\bibnamefont{Arkani-Hamed}},
  \bibinfo{author}{\bibfnamefont{S.}~\bibnamefont{Dimopoulos}},
  \bibinfo{author}{\bibfnamefont{G.~R.} \bibnamefont{Dvali}}, \bibnamefont{and}
  \bibinfo{author}{\bibfnamefont{J.}~\bibnamefont{March-Russell}},
  \bibinfo{journal}{Phys. Rev.} \textbf{\bibinfo{volume}{D65}},
  \bibinfo{pages}{024032} (\bibinfo{year}{2002}), \eprint{hep-ph/9811448}\relax
\relax
\bibitem{Dienes:1998sb}
\bibinfo{author}{\bibfnamefont{K.~R.} \bibnamefont{Dienes}},
  \bibinfo{author}{\bibfnamefont{E.}~\bibnamefont{Dudas}}, \bibnamefont{and}
  \bibinfo{author}{\bibfnamefont{T.}~\bibnamefont{Gherghetta}},
  \bibinfo{journal}{Nucl. Phys.} \textbf{\bibinfo{volume}{B557}},
  \bibinfo{pages}{25} (\bibinfo{year}{1999}), \eprint{hep-ph/9811428}\relax
\relax
\bibitem{Lukas:2000wn}
\bibinfo{author}{\bibfnamefont{A.}~\bibnamefont{Lukas}},
  \bibinfo{author}{\bibfnamefont{P.}~\bibnamefont{Ramond}},
  \bibinfo{author}{\bibfnamefont{A.}~\bibnamefont{Romanino}}, \bibnamefont{and}
  \bibinfo{author}{\bibfnamefont{G.~G.} \bibnamefont{Ross}},
  \bibinfo{journal}{Phys. Lett.} \textbf{\bibinfo{volume}{B495}},
  \bibinfo{pages}{136} (\bibinfo{year}{2000}), \eprint{hep-ph/0008049}\relax
\relax
\bibitem{Lukas:2000rg}
\bibinfo{author}{\bibfnamefont{A.}~\bibnamefont{Lukas}},
  \bibinfo{author}{\bibfnamefont{P.}~\bibnamefont{Ramond}},
  \bibinfo{author}{\bibfnamefont{A.}~\bibnamefont{Romanino}}, \bibnamefont{and}
  \bibinfo{author}{\bibfnamefont{G.~G.} \bibnamefont{Ross}},
  \bibinfo{journal}{JHEP} \textbf{\bibinfo{volume}{04}}, \bibinfo{pages}{010}
  (\bibinfo{year}{2001}), \eprint{hep-ph/0011295}\relax
\relax
\bibitem{Faraggi:1999bm}
\bibinfo{author}{\bibfnamefont{A.~E.} \bibnamefont{Faraggi}} \bibnamefont{and}
  \bibinfo{author}{\bibfnamefont{M.}~\bibnamefont{Pospelov}},
  \bibinfo{journal}{Phys. Lett.} \textbf{\bibinfo{volume}{B458}},
  \bibinfo{pages}{237} (\bibinfo{year}{1999}), \eprint{hep-ph/9901299}\relax
\relax
\bibitem{Dvali:1999cn}
\bibinfo{author}{\bibfnamefont{G.~R.} \bibnamefont{Dvali}} \bibnamefont{and}
  \bibinfo{author}{\bibfnamefont{A.~Y.} \bibnamefont{Smirnov}},
  \bibinfo{journal}{Nucl. Phys.} \textbf{\bibinfo{volume}{B563}},
  \bibinfo{pages}{63} (\bibinfo{year}{1999}), \eprint{hep-ph/9904211}\relax
\relax
\bibitem{Das:1999dx}
\bibinfo{author}{\bibfnamefont{A.~K.} \bibnamefont{Das}} \bibnamefont{and}
  \bibinfo{author}{\bibfnamefont{O.~C.~W.} \bibnamefont{Kong}},
  \bibinfo{journal}{Phys. Lett.} \textbf{\bibinfo{volume}{B470}},
  \bibinfo{pages}{149} (\bibinfo{year}{1999}), \eprint{hep-ph/9907272}\relax
\relax
\bibitem{Mohapatra:1999zd}
\bibinfo{author}{\bibfnamefont{R.~N.} \bibnamefont{Mohapatra}},
  \bibinfo{author}{\bibfnamefont{S.}~\bibnamefont{Nandi}}, \bibnamefont{and}
  \bibinfo{author}{\bibfnamefont{A.}~\bibnamefont{Perez-Lorenzana}},
  \bibinfo{journal}{Phys. Lett.} \textbf{\bibinfo{volume}{B466}},
  \bibinfo{pages}{115} (\bibinfo{year}{1999}), \eprint{hep-ph/9907520}\relax
\relax
\bibitem{Ioannisian:1999cw}
\bibinfo{author}{\bibfnamefont{A.}~\bibnamefont{Ioannisian}} \bibnamefont{and}
  \bibinfo{author}{\bibfnamefont{A.}~\bibnamefont{Pilaftsis}},
  \bibinfo{journal}{Phys. Rev.} \textbf{\bibinfo{volume}{D62}},
  \bibinfo{pages}{066001} (\bibinfo{year}{2000}), \eprint{hep-ph/9907522}\relax
\relax
\bibitem{Mohapatra:1999af}
\bibinfo{author}{\bibfnamefont{R.~N.} \bibnamefont{Mohapatra}}
  \bibnamefont{and}
  \bibinfo{author}{\bibfnamefont{A.}~\bibnamefont{Perez-Lorenzana}},
  \bibinfo{journal}{Nucl. Phys.} \textbf{\bibinfo{volume}{B576}},
  \bibinfo{pages}{466} (\bibinfo{year}{2000}), \eprint{hep-ph/9910474}\relax
\relax
\bibitem{Ioannisian:1999sw}
\bibinfo{author}{\bibfnamefont{A.}~\bibnamefont{Ioannisian}} \bibnamefont{and}
  \bibinfo{author}{\bibfnamefont{J.~W.~F.} \bibnamefont{Valle}},
  \bibinfo{journal}{Phys. Rev.} \textbf{\bibinfo{volume}{D63}},
  \bibinfo{pages}{073002} (\bibinfo{year}{2001}), \eprint{hep-ph/9911349}\relax
\relax
\bibitem{Barbieri:2000mg}
\bibinfo{author}{\bibfnamefont{R.}~\bibnamefont{Barbieri}},
  \bibinfo{author}{\bibfnamefont{P.}~\bibnamefont{Creminelli}},
  \bibnamefont{and} \bibinfo{author}{\bibfnamefont{A.}~\bibnamefont{Strumia}},
  \bibinfo{journal}{Nucl. Phys.} \textbf{\bibinfo{volume}{B585}},
  \bibinfo{pages}{28} (\bibinfo{year}{2000}), \eprint{hep-ph/0002199}\relax
\relax
\bibitem{Lukas:2000fy}
\bibinfo{author}{\bibfnamefont{A.}~\bibnamefont{Lukas}} \bibnamefont{and}
  \bibinfo{author}{\bibfnamefont{A.}~\bibnamefont{Romanino}}
  (\bibinfo{year}{2000}), \eprint{hep-ph/0004130}\relax
\relax
\bibitem{Ma:2000wp}
\bibinfo{author}{\bibfnamefont{E.}~\bibnamefont{Ma}},
  \bibinfo{author}{\bibfnamefont{M.}~\bibnamefont{Raidal}}, \bibnamefont{and}
  \bibinfo{author}{\bibfnamefont{U.}~\bibnamefont{Sarkar}},
  \bibinfo{journal}{Phys. Rev. Lett.} \textbf{\bibinfo{volume}{85}},
  \bibinfo{pages}{3769} (\bibinfo{year}{2000}), \eprint{hep-ph/0006046}\relax
\relax
\bibitem{Mohapatra:2000wn}
\bibinfo{author}{\bibfnamefont{R.~N.} \bibnamefont{Mohapatra}}
  \bibnamefont{and}
  \bibinfo{author}{\bibfnamefont{A.}~\bibnamefont{Perez-Lorenzana}},
  \bibinfo{journal}{Nucl. Phys.} \textbf{\bibinfo{volume}{B593}},
  \bibinfo{pages}{451} (\bibinfo{year}{2001}), \eprint{hep-ph/0006278}\relax
\relax
\bibitem{Ma:2000gf}
\bibinfo{author}{\bibfnamefont{E.}~\bibnamefont{Ma}},
  \bibinfo{author}{\bibfnamefont{G.}~\bibnamefont{Rajasekaran}},
  \bibnamefont{and} \bibinfo{author}{\bibfnamefont{U.}~\bibnamefont{Sarkar}},
  \bibinfo{journal}{Phys. Lett.} \textbf{\bibinfo{volume}{B495}},
  \bibinfo{pages}{363} (\bibinfo{year}{2000}), \eprint{hep-ph/0006340}\relax
\relax
\bibitem{Dienes:2000ph}
\bibinfo{author}{\bibfnamefont{K.~R.} \bibnamefont{Dienes}} \bibnamefont{and}
  \bibinfo{author}{\bibfnamefont{I.}~\bibnamefont{Sarcevic}},
  \bibinfo{journal}{Phys. Lett.} \textbf{\bibinfo{volume}{B500}},
  \bibinfo{pages}{133} (\bibinfo{year}{2001}), \eprint{hep-ph/0008144}\relax
\relax
\bibitem{Mohapatra:2000px}
\bibinfo{author}{\bibfnamefont{R.~N.} \bibnamefont{Mohapatra}},
  \bibinfo{author}{\bibfnamefont{A.}~\bibnamefont{Perez-Lorenzana}},
  \bibnamefont{and} \bibinfo{author}{\bibfnamefont{C.~A.}
  \bibnamefont{de~S~Pires}}, \bibinfo{journal}{Phys. Lett.}
  \textbf{\bibinfo{volume}{B491}}, \bibinfo{pages}{143} (\bibinfo{year}{2000}),
  \eprint{hep-ph/0008158}\relax
\relax
\bibitem{McLaughlin:2000zf}
\bibinfo{author}{\bibfnamefont{G.~C.} \bibnamefont{McLaughlin}}
  \bibnamefont{and} \bibinfo{author}{\bibfnamefont{J.~N.} \bibnamefont{Ng}},
  \bibinfo{journal}{Phys. Lett.} \textbf{\bibinfo{volume}{B493}},
  \bibinfo{pages}{88} (\bibinfo{year}{2000}), \eprint{hep-ph/0008209}\relax
\relax
\bibitem{Ioannisian:2000en}
\bibinfo{author}{\bibfnamefont{A.}~\bibnamefont{Ioannisian}} \bibnamefont{and}
  \bibinfo{author}{\bibfnamefont{A.}~\bibnamefont{Pilaftsis}},
  \bibinfo{journal}{Int. J. Mod. Phys.} \textbf{\bibinfo{volume}{A16S1C}},
  \bibinfo{pages}{931} (\bibinfo{year}{2001}), \eprint{hep-ph/0010051}\relax
\relax
\bibitem{Agashe:2000rw}
\bibinfo{author}{\bibfnamefont{K.}~\bibnamefont{Agashe}} \bibnamefont{and}
  \bibinfo{author}{\bibfnamefont{G.-H.} \bibnamefont{Wu}},
  \bibinfo{journal}{Phys. Lett.} \textbf{\bibinfo{volume}{B498}},
  \bibinfo{pages}{230} (\bibinfo{year}{2001}), \eprint{hep-ph/0010117}\relax
\relax
\bibitem{Cosme:2000ib}
\bibinfo{author}{\bibfnamefont{N.}~\bibnamefont{Cosme}} \emph{et~al.},
  \bibinfo{journal}{Phys. Rev.} \textbf{\bibinfo{volume}{D63}},
  \bibinfo{pages}{113018} (\bibinfo{year}{2001}), \eprint{hep-ph/0010192}\relax
\relax
\bibitem{Caldwell:2000zn}
\bibinfo{author}{\bibfnamefont{D.~O.} \bibnamefont{Caldwell}},
  \bibinfo{author}{\bibfnamefont{R.~N.} \bibnamefont{Mohapatra}},
  \bibnamefont{and} \bibinfo{author}{\bibfnamefont{S.~J.}
  \bibnamefont{Yellin}}, \bibinfo{journal}{Phys. Rev. Lett.}
  \textbf{\bibinfo{volume}{87}}, \bibinfo{pages}{041601}
  (\bibinfo{year}{2001}), \eprint{hep-ph/0010353}\relax
\relax
\bibitem{Abazajian:2000hw}
\bibinfo{author}{\bibfnamefont{K.}~\bibnamefont{Abazajian}},
  \bibinfo{author}{\bibfnamefont{G.~M.} \bibnamefont{Fuller}},
  \bibnamefont{and} \bibinfo{author}{\bibfnamefont{M.}~\bibnamefont{Patel}},
  \bibinfo{journal}{Phys. Rev. Lett.} \textbf{\bibinfo{volume}{90}},
  \bibinfo{pages}{061301} (\bibinfo{year}{2003}), \eprint{hep-ph/0011048}\relax
\relax
\bibitem{Caldwell:2001dj}
\bibinfo{author}{\bibfnamefont{D.~O.} \bibnamefont{Caldwell}},
  \bibinfo{author}{\bibfnamefont{R.~N.} \bibnamefont{Mohapatra}},
  \bibnamefont{and} \bibinfo{author}{\bibfnamefont{S.~J.}
  \bibnamefont{Yellin}}, \bibinfo{journal}{Phys. Rev.}
  \textbf{\bibinfo{volume}{D64}}, \bibinfo{pages}{073001}
  (\bibinfo{year}{2001}), \eprint{hep-ph/0102279}\relax
\relax
\bibitem{Lam:2001gy}
\bibinfo{author}{\bibfnamefont{C.~S.} \bibnamefont{Lam}} \bibnamefont{and}
  \bibinfo{author}{\bibfnamefont{J.~N.} \bibnamefont{Ng}},
  \bibinfo{journal}{Phys. Rev.} \textbf{\bibinfo{volume}{D64}},
  \bibinfo{pages}{113006} (\bibinfo{year}{2001}), \eprint{hep-ph/0104129}\relax
\relax
\bibitem{DeGouvea:2001mz}
\bibinfo{author}{\bibfnamefont{A.}~\bibnamefont{De~Gouvea}},
  \bibinfo{author}{\bibfnamefont{G.~F.} \bibnamefont{Giudice}},
  \bibinfo{author}{\bibfnamefont{A.}~\bibnamefont{Strumia}}, \bibnamefont{and}
  \bibinfo{author}{\bibfnamefont{K.}~\bibnamefont{Tobe}},
  \bibinfo{journal}{Nucl. Phys.} \textbf{\bibinfo{volume}{B623}},
  \bibinfo{pages}{395} (\bibinfo{year}{2002}), \eprint{hep-ph/0107156}\relax
\relax
\bibitem{Frere:2001ug}
\bibinfo{author}{\bibfnamefont{J.~M.} \bibnamefont{Frere}},
  \bibinfo{author}{\bibfnamefont{M.~V.} \bibnamefont{Libanov}},
  \bibnamefont{and} \bibinfo{author}{\bibfnamefont{S.~V.}
  \bibnamefont{Troitsky}}, \bibinfo{journal}{JHEP}
  \textbf{\bibinfo{volume}{11}}, \bibinfo{pages}{025} (\bibinfo{year}{2001}),
  \eprint{hep-ph/0110045}\relax
\relax
\bibitem{Lam:2001iy}
\bibinfo{author}{\bibfnamefont{C.~S.} \bibnamefont{Lam}},
  \bibinfo{journal}{Phys. Rev.} \textbf{\bibinfo{volume}{D65}},
  \bibinfo{pages}{053009} (\bibinfo{year}{2002}), \eprint{hep-ph/0110142}\relax
\relax
\bibitem{Davoudiasl:2002fq}
\bibinfo{author}{\bibfnamefont{H.}~\bibnamefont{Davoudiasl}},
  \bibinfo{author}{\bibfnamefont{P.}~\bibnamefont{Langacker}},
  \bibnamefont{and}
  \bibinfo{author}{\bibfnamefont{M.}~\bibnamefont{Perelstein}},
  \bibinfo{journal}{Phys. Rev.} \textbf{\bibinfo{volume}{D65}},
  \bibinfo{pages}{105015} (\bibinfo{year}{2002}), \eprint{hep-ph/0201128}\relax
\relax
\bibitem{Mohapatra:2002ug}
\bibinfo{author}{\bibfnamefont{R.~N.} \bibnamefont{Mohapatra}}
  \bibnamefont{and}
  \bibinfo{author}{\bibfnamefont{A.}~\bibnamefont{Perez-Lorenzana}},
  \bibinfo{journal}{Phys. Rev.} \textbf{\bibinfo{volume}{D67}},
  \bibinfo{pages}{075015} (\bibinfo{year}{2003}), \eprint{hep-ph/0212254}\relax
\relax
\bibitem{Kim:2002im}
\bibinfo{author}{\bibfnamefont{H.~D.} \bibnamefont{Kim}} \bibnamefont{and}
  \bibinfo{author}{\bibfnamefont{S.}~\bibnamefont{Raby}},
  \bibinfo{journal}{JHEP} \textbf{\bibinfo{volume}{01}}, \bibinfo{pages}{056}
  (\bibinfo{year}{2003}), \eprint{hep-ph/0212348}\relax
\relax
\bibitem{Cao:2003yx}
\bibinfo{author}{\bibfnamefont{Q.-H.} \bibnamefont{Cao}},
  \bibinfo{author}{\bibfnamefont{S.}~\bibnamefont{Gopalakrishna}},
  \bibnamefont{and} \bibinfo{author}{\bibfnamefont{C.~P.} \bibnamefont{Yuan}},
  \bibinfo{journal}{Phys. Rev.} \textbf{\bibinfo{volume}{D69}},
  \bibinfo{pages}{115003} (\bibinfo{year}{2004}), \eprint{hep-ph/0312339}\relax
\relax
\bibitem{Hewett:2004py}
\bibinfo{author}{\bibfnamefont{J.~L.} \bibnamefont{Hewett}},
  \bibinfo{author}{\bibfnamefont{P.}~\bibnamefont{Roy}}, \bibnamefont{and}
  \bibinfo{author}{\bibfnamefont{S.}~\bibnamefont{Roy}},
  \bibinfo{journal}{Phys. Rev.} \textbf{\bibinfo{volume}{D70}},
  \bibinfo{pages}{051903(R)} (\bibinfo{year}{2004}),
  \eprint{hep-ph/0404174}\relax
\relax
\bibitem{Cao:2004tu}
\bibinfo{author}{\bibfnamefont{Q.-H.} \bibnamefont{Cao}},
  \bibinfo{author}{\bibfnamefont{S.}~\bibnamefont{Gopalakrishna}},
  \bibnamefont{and} \bibinfo{author}{\bibfnamefont{C.~P.} \bibnamefont{Yuan}},
  \bibinfo{journal}{Phys. Rev.} \textbf{\bibinfo{volume}{D70}},
  \bibinfo{pages}{075020} (\bibinfo{year}{2004}), \eprint{hep-ph/0405220}\relax
\relax
\bibitem{Dudas:2005vn}
\bibinfo{author}{\bibfnamefont{E.}~\bibnamefont{Dudas}},
  \bibinfo{author}{\bibfnamefont{C.}~\bibnamefont{Grojean}}, \bibnamefont{and}
  \bibinfo{author}{\bibfnamefont{S.~K.} \bibnamefont{Vempati}}
  (\bibinfo{year}{2005}), \eprint{hep-ph/0511001}\relax
\relax
\bibitem{Haba:2006dz}
\bibinfo{author}{\bibfnamefont{N.}~\bibnamefont{Haba}},
  \bibinfo{author}{\bibfnamefont{A.}~\bibnamefont{Watanabe}}, \bibnamefont{and}
  \bibinfo{author}{\bibfnamefont{K.}~\bibnamefont{Yoshioka}}
  (\bibinfo{year}{2006}), \eprint{hep-ph/0603116}\relax
\relax
\bibitem{King:1998jw}
\bibinfo{author}{\bibfnamefont{S.~F.} \bibnamefont{King}},
  \bibinfo{journal}{Phys. Lett.} \textbf{\bibinfo{volume}{B439}},
  \bibinfo{pages}{350} (\bibinfo{year}{1998}), \eprint{hep-ph/9806440}\relax
\relax
\bibitem{Davidson:1998bi}
\bibinfo{author}{\bibfnamefont{S.}~\bibnamefont{Davidson}} \bibnamefont{and}
  \bibinfo{author}{\bibfnamefont{S.~F.} \bibnamefont{King}},
  \bibinfo{journal}{Phys. Lett.} \textbf{\bibinfo{volume}{B445}},
  \bibinfo{pages}{191} (\bibinfo{year}{1998}), \eprint{hep-ph/9808296}\relax
\relax
\bibitem{King:2003jb}
\bibinfo{author}{\bibfnamefont{S.~F.} \bibnamefont{King}},
  \bibinfo{journal}{Rept. Prog. Phys.} \textbf{\bibinfo{volume}{67}},
  \bibinfo{pages}{107} (\bibinfo{year}{2004}), \eprint{hep-ph/0310204}\relax
\relax
\bibitem{King:1999cm}
\bibinfo{author}{\bibfnamefont{S.~F.} \bibnamefont{King}},
  \bibinfo{journal}{Nucl. Phys.} \textbf{\bibinfo{volume}{B562}},
  \bibinfo{pages}{57} (\bibinfo{year}{1999}), \eprint{hep-ph/9904210}\relax
\relax
\bibitem{King:1999mb}
\bibinfo{author}{\bibfnamefont{S.~F.} \bibnamefont{King}},
  \bibinfo{journal}{Nucl. Phys.} \textbf{\bibinfo{volume}{B576}},
  \bibinfo{pages}{85} (\bibinfo{year}{2000}), \eprint{hep-ph/9912492}\relax
\relax
\bibitem{Haba:2006gt}
\bibinfo{author}{\bibfnamefont{N.}~\bibnamefont{Haba}}, \bibinfo{journal}{JHEP}
  \textbf{\bibinfo{volume}{05}}, \bibinfo{pages}{030} (\bibinfo{year}{2006}),
  \eprint{hep-ph/0603119}\relax
\relax
\bibitem{Arkani-Hamed:1998rs}
\bibinfo{author}{\bibfnamefont{N.}~\bibnamefont{Arkani-Hamed}},
  \bibinfo{author}{\bibfnamefont{S.}~\bibnamefont{Dimopoulos}},
  \bibnamefont{and} \bibinfo{author}{\bibfnamefont{G.~R.} \bibnamefont{Dvali}},
  \bibinfo{journal}{Phys. Lett.} \textbf{\bibinfo{volume}{B429}},
  \bibinfo{pages}{263} (\bibinfo{year}{1998}), \eprint{hep-ph/9803315}\relax
\relax
\bibitem{Antoniadis:1998ig}
\bibinfo{author}{\bibfnamefont{I.}~\bibnamefont{Antoniadis}},
  \bibinfo{author}{\bibfnamefont{N.}~\bibnamefont{Arkani-Hamed}},
  \bibinfo{author}{\bibfnamefont{S.}~\bibnamefont{Dimopoulos}},
  \bibnamefont{and} \bibinfo{author}{\bibfnamefont{G.~R.} \bibnamefont{Dvali}},
  \bibinfo{journal}{Phys. Lett.} \textbf{\bibinfo{volume}{B436}},
  \bibinfo{pages}{257} (\bibinfo{year}{1998}), \eprint{hep-ph/9804398}\relax
\relax
\bibitem{Froggatt:1978nt}
\bibinfo{author}{\bibfnamefont{C.~D.} \bibnamefont{Froggatt}} \bibnamefont{and}
  \bibinfo{author}{\bibfnamefont{H.~B.} \bibnamefont{Nielsen}},
  \bibinfo{journal}{Nucl. Phys.} \textbf{\bibinfo{volume}{B147}},
  \bibinfo{pages}{277} (\bibinfo{year}{1979})\relax
\relax
\bibitem{Ellis:1998nk}
\bibinfo{author}{\bibfnamefont{J.~R.} \bibnamefont{Ellis}},
  \bibinfo{author}{\bibfnamefont{G.~K.} \bibnamefont{Leontaris}},
  \bibinfo{author}{\bibfnamefont{S.}~\bibnamefont{Lola}}, \bibnamefont{and}
  \bibinfo{author}{\bibfnamefont{D.~V.} \bibnamefont{Nanopoulos}},
  \bibinfo{journal}{Eur. Phys. J.} \textbf{\bibinfo{volume}{C9}},
  \bibinfo{pages}{389} (\bibinfo{year}{1999}), \eprint{hep-ph/9808251}\relax
\relax
\bibitem{Lola:1999un}
\bibinfo{author}{\bibfnamefont{S.}~\bibnamefont{Lola}} \bibnamefont{and}
  \bibinfo{author}{\bibfnamefont{G.~G.} \bibnamefont{Ross}},
  \bibinfo{journal}{Nucl. Phys.} \textbf{\bibinfo{volume}{B553}},
  \bibinfo{pages}{81} (\bibinfo{year}{1999}), \eprint{hep-ph/9902283}\relax
\relax
\bibitem{Leontaris:1995be}
\bibinfo{author}{\bibfnamefont{G.~K.} \bibnamefont{Leontaris}},
  \bibinfo{author}{\bibfnamefont{S.}~\bibnamefont{Lola}}, \bibnamefont{and}
  \bibinfo{author}{\bibfnamefont{G.~G.} \bibnamefont{Ross}},
  \bibinfo{journal}{Nucl. Phys.} \textbf{\bibinfo{volume}{B454}},
  \bibinfo{pages}{25} (\bibinfo{year}{1995}), \eprint{hep-ph/9505402}\relax
\relax
\bibitem{Dreiner:1994ra}
\bibinfo{author}{\bibfnamefont{H.~K.} \bibnamefont{Dreiner}},
  \bibinfo{author}{\bibfnamefont{G.~K.} \bibnamefont{Leontaris}},
  \bibinfo{author}{\bibfnamefont{S.}~\bibnamefont{Lola}},
  \bibinfo{author}{\bibfnamefont{G.~G.} \bibnamefont{Ross}}, \bibnamefont{and}
  \bibinfo{author}{\bibfnamefont{C.}~\bibnamefont{Scheich}},
  \bibinfo{journal}{Nucl. Phys.} \textbf{\bibinfo{volume}{B436}},
  \bibinfo{pages}{461} (\bibinfo{year}{1995}), \eprint{hep-ph/9409369}\relax
\relax
\bibitem{Elwood:1998kf}
\bibinfo{author}{\bibfnamefont{J.~K.} \bibnamefont{Elwood}},
  \bibinfo{author}{\bibfnamefont{N.}~\bibnamefont{Irges}}, \bibnamefont{and}
  \bibinfo{author}{\bibfnamefont{P.}~\bibnamefont{Ramond}},
  \bibinfo{journal}{Phys. Rev. Lett.} \textbf{\bibinfo{volume}{81}},
  \bibinfo{pages}{5064} (\bibinfo{year}{1998}), \eprint{hep-ph/9807228}\relax
\relax
\bibitem{Irges:1998ax}
\bibinfo{author}{\bibfnamefont{N.}~\bibnamefont{Irges}},
  \bibinfo{author}{\bibfnamefont{S.}~\bibnamefont{Lavignac}}, \bibnamefont{and}
  \bibinfo{author}{\bibfnamefont{P.}~\bibnamefont{Ramond}},
  \bibinfo{journal}{Phys. Rev.} \textbf{\bibinfo{volume}{D58}},
  \bibinfo{pages}{035003} (\bibinfo{year}{1998}), \eprint{hep-ph/9802334}\relax
\relax
\bibitem{Altarelli:1998sr}
\bibinfo{author}{\bibfnamefont{G.}~\bibnamefont{Altarelli}} \bibnamefont{and}
  \bibinfo{author}{\bibfnamefont{F.}~\bibnamefont{Feruglio}},
  \bibinfo{journal}{JHEP} \textbf{\bibinfo{volume}{11}}, \bibinfo{pages}{021}
  (\bibinfo{year}{1998}), \eprint{hep-ph/9809596}\relax
\relax
\bibitem{Altarelli:1998ns}
\bibinfo{author}{\bibfnamefont{G.}~\bibnamefont{Altarelli}} \bibnamefont{and}
  \bibinfo{author}{\bibfnamefont{F.}~\bibnamefont{Feruglio}},
  \bibinfo{journal}{Phys. Lett.} \textbf{\bibinfo{volume}{B451}},
  \bibinfo{pages}{388} (\bibinfo{year}{1999}), \eprint{hep-ph/9812475}\relax
\relax
\bibitem{Altarelli:1999wi}
\bibinfo{author}{\bibfnamefont{G.}~\bibnamefont{Altarelli}} \bibnamefont{and}
  \bibinfo{author}{\bibfnamefont{F.}~\bibnamefont{Feruglio}}
  (\bibinfo{year}{1999}), \eprint{hep-ph/9905536}\relax
\relax
\bibitem{Wilczek:1982rv}
\bibinfo{author}{\bibfnamefont{F.}~\bibnamefont{Wilczek}},
  \bibinfo{journal}{Phys. Rev. Lett.} \textbf{\bibinfo{volume}{49}},
  \bibinfo{pages}{1549} (\bibinfo{year}{1982})\relax
\relax
\bibitem{Reiss:1982sq}
\bibinfo{author}{\bibfnamefont{D.~B.} \bibnamefont{Reiss}},
  \bibinfo{journal}{Phys. Lett.} \textbf{\bibinfo{volume}{B115}},
  \bibinfo{pages}{217} (\bibinfo{year}{1982})\relax
\relax
\bibitem{Gelmini:1982zz}
\bibinfo{author}{\bibfnamefont{G.~B.} \bibnamefont{Gelmini}},
  \bibinfo{author}{\bibfnamefont{S.}~\bibnamefont{Nussinov}}, \bibnamefont{and}
  \bibinfo{author}{\bibfnamefont{T.}~\bibnamefont{Yanagida}},
  \bibinfo{journal}{Nucl. Phys.} \textbf{\bibinfo{volume}{B219}},
  \bibinfo{pages}{31} (\bibinfo{year}{1983})\relax
\relax
\bibitem{Feng:1997tn}
\bibinfo{author}{\bibfnamefont{J.~L.} \bibnamefont{Feng}},
  \bibinfo{author}{\bibfnamefont{T.}~\bibnamefont{Moroi}},
  \bibinfo{author}{\bibfnamefont{H.}~\bibnamefont{Murayama}}, \bibnamefont{and}
  \bibinfo{author}{\bibfnamefont{E.}~\bibnamefont{Schnapka}},
  \bibinfo{journal}{Phys. Rev.} \textbf{\bibinfo{volume}{D57}},
  \bibinfo{pages}{5875} (\bibinfo{year}{1998})\relax
\relax
\bibitem{Ammar:2001gi}
\bibinfo{author}{\bibfnamefont{R.}~\bibnamefont{Ammar}} \emph{et~al.}
  (\bibinfo{collaboration}{CLEO}), \bibinfo{journal}{Phys. Rev. Lett.}
  \textbf{\bibinfo{volume}{87}}, \bibinfo{pages}{271801}
  (\bibinfo{year}{2001}), \eprint{hep-ex/0106038}\relax
\relax
\bibitem{Diaz-Cruz:2004sp}
\bibinfo{author}{\bibfnamefont{J.~L.} \bibnamefont{Diaz-Cruz}}
  \bibnamefont{and} \bibinfo{author}{\bibfnamefont{C.~E.}
  \bibnamefont{Pagliarone}}  (\bibinfo{year}{2004}),
  \eprint{hep-ph/0412329}\relax
\relax
\end{mcbibliography}
\end{document}